\documentclass[aps,prb,twocolumn,superscriptaddress]{revtex4}

\usepackage{amsmath}    
\usepackage{graphicx}  
\usepackage{natbib}  
\usepackage{hyperref}
\usepackage{epstopdf}
\usepackage{subfigure}
\usepackage{pstricks}
\usepackage{color}
\usepackage{mathrsfs}
\usepackage{feynmp}
\usepackage{fixmath}
\newcommand{\nn}{\nonumber \\}

\newcommand{\mbd}{\mathbold}

\begin{document}

 \title{\bf Andreev reflection in 2D relativistic materials with realistic tunneling transparency in normal-metal-superconductor junctions}

       \author{Yung-Yeh, Chang}
       \email{cdshjtr@gmail.com}
       \affiliation{Department of Electrophysics, National Chiao Tung University\\
1001 University Street, Hsinchu, Taiwan, R.O.C.}
\author{Chung-Yu, Mou}
\email{mou@phys.nthu.edu.tw}
\affiliation{Department of Physics, National Tsing Hua University, Hsinchu 30043, Taiwan, 300, R.O.C.}
\affiliation{Institute of Physics, Academia Sinica, Nankang, Taiwan, R.O.C.}
\affiliation{Physics Division, National Center for Theoretical Sciences, P. O. Box
2-131, Hsinchu, Taiwan, R.O.C.}
       \author{Chung-Hou, Chung}
       \email{chung@mail.nctu.edu.tw}
        \affiliation{Department of Electrophysics, National Chiao Tung University\\
1001 University Street, Hsinchu, Taiwan, R.O.C.}
\affiliation{Physics Division, National Center for Theoretical Sciences, P. O. Box
2-131, Hsinchu, Taiwan, R.O.C.}

\begin{abstract} 
The Andreev conductance across $2d$ normal metal (N)/superconductor (SC) junctions with relativistic Dirac spectrum is investigated theoretically in the Blonder-­Tinkham-­Klapwijk formalism. It is shown that for relativistic materials, due to the Klein tunneling instead of impurity potentials, the local strain in the junction is the key factor that determines the transparency of the junction. The local strain is shown to generate an effective Dirac $\delta$-gauge field. A remarkable suppression of the conductance are observed as the strength of the gauge field increases. The behaviors of the conductance are in well agreement with the results obtained in the case of $1d$ N/SC junction. We also study the Andreev reflection in a topological material near the chiral­-to-helical phase transition in the presence of a local strain. The N side of the N/SC junction is modeled by the doped Kane­-Mele (KM) model. The SC region is a doped correlated KM $t$-$­J$ (KMtJ) model, which has been shown to feature $d+id^{\prime}$-­wave spin-singlet pairing. With increasing intrinsic spin­-orbit (SO) coupling, the doped KMtJ system undergoes a topological phase transition from the chiral $d$­-wave superconductivity to the spin­-Chern superconducting phase with helical Majorana fermions at edges. We explore the Andreev conductance at the two inequivalent Dirac points, respectively and predict the distinctive behaviors for the Andreev conductance across the topological phase transition. Relevance of our results for the adatom­-doped graphene is discussed. 
\end{abstract}

  \maketitle
  \section{Introduction}
  The Andreev reflection (AR) is an electron-hole conversion process taking place between a normal metal and a superconductor with the excitation energy of the incident electrons is lower than the superconducting gap energy, which was originally proposed by A. F. Andreev in 1964 [\onlinecite{Andreev1964}]. This conventional Andreev reflection is also called the Andreev retro-reflection since the reflected holes will retrace the path of the incident electrons. A recent theoretical observation on the Andreev reflection at the interface between a sheet of pure graphene and a sheet of thin film graphene s-wave superconductor induced by proximity effect unvieled another Andreev reflection process, known as the specular Andreev reflection [\onlinecite{Beenakker2006a}, \onlinecite{Beenakker2008RMP}]. This process takes place when an incident electron with its excitation energy greater than the Fermi energy but lower than the superconducting gap in the conduction band is converted into a hole in the valence band associated with a reverse on the velocity parallel to the N/SC interface. Moreover, at the energy higher than the Fermi energy the Andreev conductance is dominated by the Andreev specular reflection. After the discovery of the Andreev specular reflection in graphene superconductor junction [\onlinecite{Beenakker2006a}, \onlinecite{Beenakker2008RMP}], the Andreev specular reflection is also predicted to happen in a 2$d$ semiconductor-superconductor junction with finite Rashba spin-orbit (SO) coupling [\onlinecite{SAR-semicond}]. In addition, various theoretical works concerning the Andreev reflection have been done on a variety of N/SC junctions with underlying honeycomb lattice structure such as graphene/$d+id^\prime$-wave superconductor [\onlinecite{Jiang2008}], bilayer-graphene/s-wave superconductor [\onlinecite{Ludwig2007BLGAR}] and the topological materials [\onlinecite{Majidi2014a}, \onlinecite{Majidi2015}]. These works all relies on the assumption of a clean and smooth N/SC interface. 
 Recently, the AR on a graphene-based superconducting junction has been experimentally realized in Ref. \onlinecite{Efetov2015} and \onlinecite{Sahu2016}, the Andreev conductance across a low disorder van der Waals interface formed between bilayer graphene and superconducting NbSe$_{2}$ has been measured. They found the conductance across the N/SC junction was suppressed when the Fermi level across the junction was tuned to lie within  the superconducting gap, which gave a solid evidence on the transition between Andreev retro-reflection (intra-band process) to Andreev specular reflection (inter-band process) on a graphene honeycomb lattice. Moreover, due to the limitation on the fabrication technology of transparent N/SC junctions, a finite tunneling transparency on the N/SC junction has been also measured.
  % finite barrier on the graphene-based/NbSe$_{2}$ superconductor junctions\cite{Efetov2015, Sahu2016} has been measured experimentally, 
 
  %The experimental limitation on fabricating a transparent N/SC junctions strongly suggests
The above limitations strongly suggest that the previous theoretical calculations on the Andreev conductance under the assumption of a transparent N/SC junction are not sufficient to account for the most recent experimental observations. In this paper, we try to simulate realistic N/SC junction with different degree of tunneling transparency. 
To do so, we first realize that for relatisvistic materials, the impurity potentials in the junction do not suppress the current. Instead, due to the phenomena associated with the Klein paradox, the impurity potentials tend to enhance the tunneling current [\onlinecite{Klein2006}]. Therefore, unlike the conventional Blonder-­Tinkham-­Klapwijk (BTK) formalism [\onlinecite{BTK1982}] in which the transparency of the tunneling is simulated by the impurity potential across
the junction, it is not feasible to control the transparency for the N/SC junctions made by relativistic materials via impurity potentials.

In this work, we show that the local strain in the junction is the key factor that determines the transparency of the junction made by relativistic materials. In particular, we describe how to impose different degrees of transparency by adding a narrow homogeneous local strain parallel to the N/SC interface [\onlinecite{CastroNeto2009RMP}, \onlinecite{Castro2009strain}] for a system with underlying honeycomb lattice. It is shown that  a $\delta$-gauge field on the N/SC interface arises and can be used to investitage the Andreev reflection in various degrees of tunneling transparency. We choose the graphene normal metal/$d+id^{\prime}$-wave superconductor N/SC junction as an example and compute the Andreev conductance via the BTK formalism [\onlinecite{BTK1982}] in the presence of a $\delta$-gauge potential. As the strength of the $\delta$-barrier is varied, the behaviors of Andreev conductance obtained via our approach are in good agreement with the results obtained in the case of 1D normal metal/$s$-wave N/SC junction in Ref. \onlinecite{BTK1982}.  Next, we apply our theory to the case of a $2d$ topological quantum spin Hall insulator. Due to the broken valley-degeneracy [\onlinecite{Yung2015}, \onlinecite{Mou-duality}], the Andreev conductance from the incident electrons near the two Dirac points exhibits entirely different behaviors. We provide experimental profiles of the Andreev conductance within certain parameter regimes for future experiments. 
  %At the end of this article, we propose that dopping adatoms in pure-graphene could be a possible experimental realization for the topological quantum spin-Hall insulator.
  
\section{AR across a graphene-$d+id^\prime$-wave N/SC junction in the presence of a $\delta$-gauge field}
%Motivated by the early experimental results, we investigate the Andreev reflection across a variety of normal metal-superconducting (N/SC) junctions with an underlying honeycomb lattice via BTK formalism.
 %have been widely discussed in many literatures by assuming an "ideal" N/SC junction, namely  no impurities or lattice mismatch. However, due to the limitation in the fabrication of transparent N/SC junctions, the assumptions of  made in those previous observations.
 In this section, we start by briefly review the electron motions in a sheet of graphene monolayer,  within the tight-binding formalism in Sec. \ref{sec:graphene-review}. The configurations of the lattice structure for a single-layered graphene sheet is schematically illustrated in the left region in Fig. \ref{fig:localstrain}-(a). The derivations of the effective $\delta$-gauge field via applying a narrow homogeneous local strain on graphene will be given in Sec. \ref{sec:deltapotential}. Finally, we apply the BTK formalism [\onlinecite{BTK1982}] to compute the Andreev conductance across a graphene normal metal/$d+id^\prime$-wave N/SC junction in the presence of a $\delta$-gauge field. The results will be illustrated in Sec. \ref{sec:AR-d+id}.
\subsection{Electronic properties in graphene monolayer}
\label{sec:graphene-review}
The electronic motion in an uniform undoped graphene monolayer is often formulated by the nearest-neighbor tight-binding model [\onlinecite{CastroNeto2009RMP}, \onlinecite{WallacePRgraphite}]:
\begin{align}
	H_0 = & \, -\, t \, \sum_{\langle i,j\rangle } \, \left(  c^{\dagger}_{A, i} c_{B,j} + h.c.\right) \nn
	 = &\,  \, \sum_{\mbd{k} \in BZ} \left(\, f (\mbd{k}) c^{\dagger}_{A,\mbd{k}}c_{B,\mbd{k}} + h.c. \right),
	 \label{eq:GrapheneTB}
\end{align}
where $c_{\alpha,i}(c_{\alpha,i}^{\dagger})$ annihilates (creates) an electron on the $\alpha \in \lbrace A,~B\rbrace$ sublattice in the $i$-th unit cell while the electron operators in the momentum space is simply given by the Fourier transform of $c_{\alpha,i}$ : $c_{\alpha,\mbd{k}} = \left(1/\sqrt{N_s}\right) \, \sum_{i} \, e^{i \mbd{k} \cdot \mbd{R}_i} \, c_{\alpha,i}$ with $\mbd{R}_{i}$ being the position vector of the $i$-th unit cell and $N_s$ being denoted the total number of the unit cells. Nearest-neighbor lattice vectors are $\mbd{\delta}_{1, 2,3}$ with unit length $a$ as shown in Fig. \ref{fig:localstrain}-(a). Here, we set $a = 1$ in what follows. The constant prefactor $t$ represents the hopping strength between two nearest-neighbor electrons for an uniform graphene monolayer. $f(\mbd{k}) \equiv -t \sum_{i=1}^{3} \, e^{i \mbd{k} \cdot \mbd{\delta}_i} $ is a $\mbd{k}$-dependent function which characterizes the band structures. The undoped single-layered graphene based on the tight-binding Hamiltonian in Eq. (\ref{eq:GrapheneTB}) features the well-known Dirac band structure with linear spectrum on the Dirac points as shown in Fig. \ref{fig:localstrain}-(b):
\begin{align}
	\mbd{K}_+ = \left(  0,~-\frac{4\pi}{3\sqrt{3} }		\right),~~\mbd{K}_- = \left(  0,~\frac{4\pi}{3\sqrt{3}}		\right).
\end{align}
 The linear dispersion is governed by the linearized Hamiltonian around the Dirac points, which is given by $H = H_+ + H_- = \sum_{\mbd{q},\tau = \pm} \, \Psi^{\dagger}_{\tau} (\mbd{q}) \mathcal{H}_{\tau} (\mbd{q}) \Psi_{\tau}(\mbd{q})$ subjected to the condition $|\mbd{q} | \ll 1$. In the momentum space, the $2 \times 2$ Dirac Hamiltonians around the Dirac points take the form of 
\begin{align}
\mathcal{H}_{+}(\mbd{q}) & = \frac{3t}{2} \, \begin{pmatrix}
0 & iq_x -q_y \\
-iq_x-q_y & 0
\end{pmatrix}
 \nn[5pt]
&= \hbar v_F \left(\pi^{y *}q_x - \pi^x q_y \right) 
\label{eq:KpH}
\end{align}
and
\begin{align}
\mathcal{H}_{-}(\mbd{q}) & = \frac{3t}{2} \, \begin{pmatrix}
0 & iq_x +q_y \\
-iq_x+q_y & 0
\end{pmatrix} \nn [5pt]
&= -\hbar v_F \left(\pi^y q_x - \pi^x q_y \right),
  \label{eq:KmH}
\end{align}
which acts on a two-dimensional spinor $\Psi_\tau(\mbd{q}) = \left( c_{A\tau}(\mbd{q}),~ c_{B\tau}(\mbd{q})\right)^T$. The valley indices $\tau = \pm$ refer to the electronic states $\Psi_\tau (\mbd{q})$ near $\mbd{K_\pm}$. $v_F \equiv  3t/2\hbar$ is defined as the Fermi velocity for the tight-binding model of graphene. Here, $\pi^{x,y,z}$ denote the Pauli matrices:
\begin{align}
	\pi^x = \begin{pmatrix}
			 0 & 1 \\
			 1 & 0
	\end{pmatrix}, ~~~\pi^y =  \begin{pmatrix}
			 0 & -i \\
			 i & 0
	\end{pmatrix},~~\pi^z = \begin{pmatrix}
			 1 & 0 \\
			 0& -1
	\end{pmatrix},
	\label{eq:Pauil-sublattice}
\end{align}
which are used to label the sublattices. We also defined a $2\times 2$ unit matrix, 
\begin{align}
	\ \pi^0 = \begin{pmatrix}
			 1 & 0\\
			 0 & 1
	\end{pmatrix}
	\label{eq:tau0-sublattice}
\end{align}
for later use, which is also for the sublattices.
\subsection{The effective $\delta$-gauge field}
\label{sec:deltapotential}
\begin{figure}
  \centering
  \includegraphics[scale=0.42]{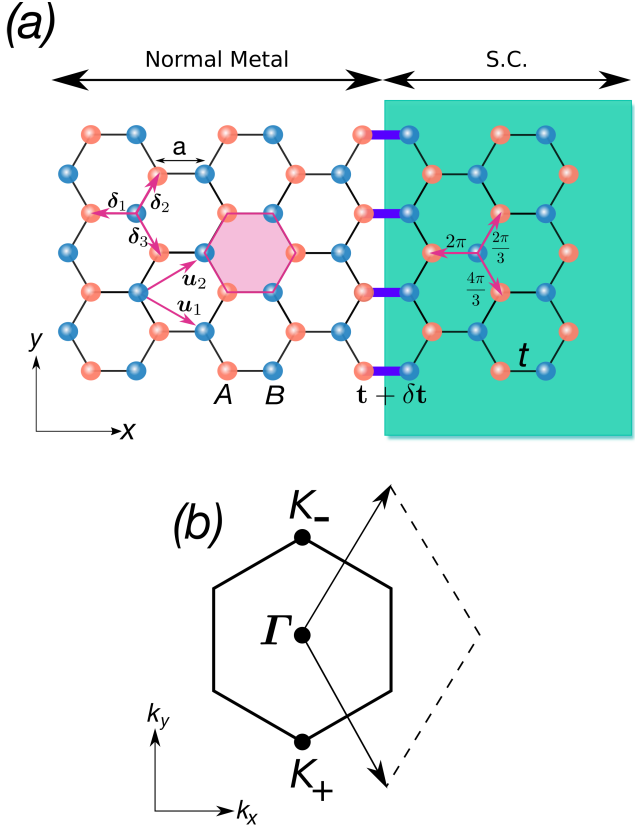}
  \caption{We consider a N/SC junction which is composed of normal metal (N) occupied the left side connecting to the superconducting (SC) region at the right ( green shaded area) with underlying honeycomb lattice structure as depicted in (a). The unit length is chosen as the nearest-neighbor lattice spacing $a =1$ throughout this article. The three phases for the bond-dependent $d+id^\prime$-pairing are defined as $\varphi_{a=1,2,3} = 2 (a-1) \pi/3$, as shown in the SC region in (a). The thicker purple horizontal bonds which locate at the N/SC interface represent the modified hopping strength $t+\delta t$ due to the local strain. The first Brillouin zone is shown in (b).}
  \label{fig:localstrain}
\end{figure}

One possible way to introduce disorders in graphene is to change the bond spacing between two different sites via applying a local strain, which effectively varies the hopping strength $t$ as in Eq. (\ref{eq:GrapheneTB}) [\onlinecite{CastroNeto2009RMP}, \onlinecite{Castro2009strain}]. To account for the effect of the local strain, we may change the hopping strength as $t\rightarrow t + \delta t(\mbd{R}_i, \,\mbd{\delta}_a)$ in the tight-binding Hamiltonian in Eq. (\ref{eq:GrapheneTB}). The magnitude of $\delta t (\mbd{R}_i, \,\mbd{\delta}_a)$ can be in general bond-dependent and spatially inhomogeneous over one bond-spacing. For simplicity, $ \delta t (\mbd{R}_i, \,\mbd{\delta}_a)$ here is assumed to be uniform over the bond spacing, thus it will not acquire Fourier components in the Fourier transformation. Under these assumptions, the linearized Hamiltonian for the change of the hopping strength around $\mbd{K}_+$-valley takes the form
\begin{alignat}{2}
 \delta H_{+} &\ = && - \int \, d\mbd{r} \, \left[A(\mbd{r}) c_{A+}^{\dagger}(\mbd{r}) c_{B+}(\mbd{r}) +A^{*}({\bf r}) c_{B+}^{\dagger}(\mbd{r}) c_{A+}(\mbd{r}) \right]\nn[5pt]
  &\ = &&  -\int \, d \mbd{r} \,  \Psi^{\dagger}_{+}(\mbd{r}) \, \begin{pmatrix}
  0 & i\mathcal{A}_x -\mathcal{A}_y \\
 - i\mathcal{A}_x -\mathcal{A}_y & 0
\end{pmatrix}  \, \Psi_{+}(\mbd{r}), \nn[5pt]
 &\ = &&  -\int \, d\mbd{r}\,  \Psi^{\dagger}_{+}(\mbd{r}) \left( \pi^{y *} \mathcal{A}_x - \pi^x \mathcal{A}_y \right)\, \Psi_{+}(\mbd{r}),
 \label{eq:strainK}
  \end{alignat}
  where $\Psi_{\tau}(\mbd{r})  =  \left( c_{A \tau}(\mbd{r}),~c_{B \tau} (\mbd{r}) \right)^{T}$ denotes the field operator for the $\mbd{K}_\tau$-valley while 
\begin{align}
A(\mbd{r}) \equiv  \sum_{a=1}^{3} \, \delta t(\mbd{r}, \mbd{\delta}_{a}) e^{i \mbd{K} \cdot \mbd{\delta}_{a}} \equiv  i\mathcal{A}_{x}(\mbd{r}) - \mathcal{A}_{y}(\mbd{r})
\label{eq:Aofr}
\end{align}
is a complex function. Here, $\mathcal{A}_x$ and $A_y$ are real functions. If we assume that the affect of the local strain only extends over one lattice spacing and only influences on the horizontal bonds along the $y$-direction, as shown in Fig. \ref{fig:localstrain} (a). The complex function $A(\mbd{r})$ in Eq. (\ref{eq:Aofr}) can be reduced to the form of a $\delta$-function. In unit of $\hbar = v_F = 1$, $A(\mbd{r})$ simply takes the form
\begin{align}
	A(\mbd{r}) & = \, \delta t (\mbd{r})  = \frac{\delta t}{t} \delta(x).
	\label{eq:vectorA}
\end{align}
Consequently, only the real part of $A(\mbd{r})$ survives; the imaginary part vanishes: $\mathcal{A}_x = 0;~~\mathcal{A}_y = - \frac{\delta t}{t} \delta(x)$

The linearized Hamiltonian for change of the hopping amplitude near $\mbd{K}_-$-valley is related to the one for the $\mbd{K}_+$-valley by time-reversal transformation $\mathcal{T}$, i.e. $\delta H_- = \mathcal{T} \, \delta H_+ \mathcal{T}^{-1}$ : 
 \begin{align}
 	\delta H_- = -\int \, d^{2}r \,  \Psi^{\dagger}_{-}({\bf r}) \left( \pi^y \mathcal{A}_x - \pi^x \mathcal{A}_y \right)\, \Psi_{-}({\bf r}).
 	\label{eq:strainKprime}  
 \end{align}
Combining Eq. {\ref{eq:KpH}} and Eq. (\ref{eq:KmH}), Eq. (\ref{eq:strainK}) and Eq. (\ref{eq:strainKprime}), the linearized Hamiltonian in the presence of a homogeneous local strain in real space is given by
\begin{widetext}
\begin{align}
	H =  \int \, d^2r \, \Psi^{\dagger}(\mbd{r}) \begin{pmatrix}
	\pi^{y *} \left( \hat{q}_x - \mathcal{A}_x\right) - \pi^x \left(\hat{q}_y -\mathcal{A}_y\right) & 0 \\[6pt]
	0 & -\left[
	 \pi^y \left( \hat{q}_x + \mathcal{A}_x \right) - \pi^x \left(\hat{q}_y + \mathcal{A}_y\right) 
	\right] 
	\end{pmatrix}  \Psi(\mbd{r}),
	\label{eq:Hrealspace}
\end{align}
\end{widetext}
 where $\Psi(\mbd{r}) \equiv \left( \Psi_+(\mbd{r}),~\Psi_-(\mbd{r})  \right)^T$ and the momentum operator $\hat{q}_i \equiv -i \partial_i$. The fact that the reverse of sign in the terms containing the complex vector $\vec{\mathcal{A} }= \left( \mathcal{A}_{x} , ~\mathcal{A}_{y}  \right)$ for different valleys in Eq. (\ref{eq:Hrealspace}) implies that $\vec{\mathcal{A}} $ can be viewed as a gauge field [\onlinecite{CastroNeto2009RMP}, \onlinecite{Castro2009strain}]. Combining Eq. (\ref{eq:Aofr}), Eq. (\ref{eq:vectorA}) and Eq. (\ref{eq:Hrealspace}), it is clear that the effect of a homogeneous local strain in the distance over one horizontal bond on graphene can be simply regarded as an effective $\delta$-gauge field of a series of localized impurities along the $y$-direction, which couples the electrons from the sublattices $A$ and $B$. 
% Apply a local strain on the interface of the N-S junction :
%\begin{align}
   %f'_{{\bf k}} = f_{{\bf k}}+\delta f_{{k\bf }}=-\sum_{\delta_{i}} (t+\delta t(\delta_{i})) e^{i\vec{k} \cdot \vec{\delta_{i}}}.
%\end{align}
\subsection{Andreev conductance across a graphene ${\it d_{x^2 - y^2} + id^{\prime}_{xy}}$-wave N/SC junction}
\label{sec:AR-d+id}
In this section, we dedicate our efforts to investigating the Andreev reflection through a graphene normal metal/$d+id^{\prime}$-wave spin-singlet superconductor N/SC junction with an effective  $\delta$-gauge field laying on the N/SC interface via BTK formalism.

Unlike the well-known case of superconductivity in graphene via promixity to a superconducting electrode [\onlinecite{heersche2007grapheneproximity}], the $d+id^{\prime}$-wave spin-singlet superconducting order is induced in graphene at finite doping by on-site electron-electron Coulomb repulsion [\onlinecite{Schaffer2007PRB}, \onlinecite{SchafferRev-chiral-dwave}, \onlinecite{LeHur-dwave-honeycomb}]. Due to the $C_{6}$ point group symmetry of the underlying honeycomb lattice, the $d+id^{\prime}$-wave superconducting order in the $\mbd{k}$-space takes the form
\begin{align}
	 \Delta_{\mbd{k}}=\sum_{a =1}^{3} \, \Delta_{\mbd{\delta}_a} e^{i\mbd{k} \cdot \mbd{\delta}_a}
\end{align}
 with the bond-dependent order parameter $\Delta_{\mbd{\delta}_a} =\Delta_0 e^{i\varphi_{a}}$ with $\varphi_{a} = 2 (a-1) \pi/3$ [\onlinecite{Schaffer2007PRB}, \onlinecite{SchafferRev-chiral-dwave}, \onlinecite{LeHur-dwave-honeycomb}].
 
 As depicted in Fig. \ref{fig:localstrain}-(a), the N/SC junction being considered is composed of a sheet of graphene normal metal (N) occupies the region of $-\infty < x < 0$ connecting to a $2d$ superconducting (SC) thin film which occupies $0 <x < \infty$ with a sharp N/SC interface in between (i.e. at the position of $x = 0$), which implies that the translational invariance along the $x$-direction is broken. We assume that the N/SC junction is homogeneous and infinitely extended in the $y$-direction, therefore the translational symmetry is preserved in $y$. The sharp N/SC junction signifies that the bulk value of the superconducting pairing amplitude denoted as $\Delta_0$ is reached at a negligibly small distance from the interface, which can be achieved via adjusting the doping or gate voltage in the SC region [\onlinecite{Beenakker2006a}, \onlinecite{Beenakker2008RMP}, \onlinecite{Ludwig2007BLGAR}]. 

Due to the valley and spin degeneracy, the electronic motions can be described by two sets of decoupled Dirac-Bogoliubov-de Gennes (DBdG) equations [\onlinecite{TinkhamBook}, \onlinecite{deGennesBook}] for the $\mbd{K}_+$ and $\mbd{K}_-$-valley, each containing four equations. Thus, it suffices to consider the set for $K_-$-valley only:
%$H_{DBdG} = \sum_{\mbd{q},\tau = \pm} \Psi_{\tau}^{\dagger}(\mbd{q}) \mathcal{M}_{\tau}(\mbd{q}) \Psi_{\tau}(\mbd{q})$
%\begin{align}
   % \mathcal{M}_{\tau}(\mbd{q} ) =  \begin{pmatrix}
      %    \mathcal{H}_{\tau}(\mbd{q} )  - \mu & \Theta(x) \Delta_{\tau}(\mbd{q} ) \\[4pt]
         %\Theta(x)  \Delta_{\tau}(\mbd{q} )^{\dagger} & \mu-\mathcal{H}_{\tau}(\mbd{q} )
    %\end{pmatrix} ,
%\end{align}
\vspace{3pt}	
		\begin{align}
			 \begin{pmatrix}
          \mathcal{H}_{-}(\mbd{q} )  - \mu & \Theta(x) \Delta_{-}(\mbd{q} ) \\[4pt]
         \Theta(x)  \Delta_{-}(\mbd{q} )^{\dagger} & \mu  - \mathcal{H}_{-}(\mbd{q} )
    \end{pmatrix}  \begin{pmatrix}
    	u \\[4pt]
    	v
    \end{pmatrix} =\frac{ \epsilon_{\mbd{q}} }{t}  \begin{pmatrix}
    	u \\[4pt]
    	v
    \end{pmatrix}, 
    \label{eq:DBdG-eq}
		\end{align}
where 
	\begin{align}
		H_{-} (\mbd{q} ) = -(  \pi^{x} q_{y} - \pi^{y} q_{x})-(U_0/t) \pi^0 \Theta(x)
	\end{align}
is the $2 \times 2 $ linearized single-particle Hamiltonian of graphene at the $\mbd{K}_-$-valley and $\mu$ denotes the chemical potential. Here, $u = (u_{A\uparrow},~ u_{B\uparrow})$ and $v = (v_{A \downarrow},~v_{B\downarrow})$ are the two component $\mbd{q}$-dependent wavefunctions for the electron (electron-like) and hole (hole-like) excitations at the excitation energy $\epsilon_\mbd{q} > 0$ which is measured relative the chemical potential $\mu$.
% and we defined four-dimensional Nambu spinor at the $\mbd{K}_-$-valley as $\Psi_{\tau}(\mbd{q} )=(c_{A\uparrow\tau}(\mbd{q}) , c_{ B\uparrow \tau} (\mbd{q} ) , c^{\dagger}_{ A\downarrow \tau} (\mbd{q} ) , c_{B \downarrow \tau}^{\dagger} (\mbd{q} ))^{T}$. 
 Here, we introduce an unit step electrostatic potential
		\begin{align}
			U_0 \Theta(x) = \begin{cases}
				U_0,~~x \geq 0\\
				0, ~~x < 0
			\end{cases}
		\end{align}
		to the SC region, where $U_0$ can be tuned independently through doping or gate voltage. To justify the assumption of sharp N/SC junction as mentioned previously, the energy scales must satisfy $U_0 \gg t \gg \mu,~\Delta_0$ such that the Fermi wavelength $\lambda_F^{\prime} = 2 \pi \hbar v_F/ (\mu + U_0)$ in SC is much shorter than that in N where $\lambda_F = 2 \pi \hbar v_F/ \mu$ [\onlinecite{Beenakker2006a}, \onlinecite{Beenakker2008RMP}, \onlinecite{Ludwig2007BLGAR}]. The linearized superconducting pairing matrix $\Delta_-(\mbd{q})$ with $d_{x^2 - y^2} + id^{\prime}_{xy}$ pairing symmetry at the $\mbd{K}_-$-valley is given by
		\begin{align}
			\Delta_- (\mbd{q}) = \frac{\Delta_0}{t} \, \begin{pmatrix}
				0 & -\frac{3}{2} (iq_x - q_y) \\[4pt]
				3  & 0
			\end{pmatrix},
			\label{eq:d+id-linearized}
		\end{align}
which is related to the one for the $K_+$-valley by $\Delta_+ (\mbd{q}) = \Delta_{-}^{T} (-\mbd{q})$. Apparently, the $d_{x^2-y^2} + id_{xy}$-wave superconducting pairing at low energy features the $s$- and $p_x + i p_y$-pairing symmetry [\onlinecite{Jiang2008}, \onlinecite{Uchoa2007}].

To study the Andreev reflection across a N/SC junction, we may image that an incident electron  comes from $x= -\infty$ toward the N/SC junction and scatters by the potential at the interface. While scattering with the potential, electrons may reflect back to the N region either as normal electrons or holes, or may tunnel through the barrier into the SC region as Dirac-Bogoliubov quasiparticles. For convenience, electrons are assumed to go through elastic scattering processes at the interface. Hence, all the scattering basis for the incident, reflected and transmitted states inside the N and SC region, which can be solved via Eq. (\ref{eq:DBdG-eq}) respectively, are characterized by the same excitation energy $\epsilon$. 
% are needed to be prepared for the future computations of the reflection and transmission coefficients.%Furthermore, the valley and spin degeneracy enable us to only consider the incident states to be spin up and from the $K_-$-valley.

 The real space eigenfucntions of Eq. (\ref{eq:DBdG-eq}) in general take the form of plane-wave solutions, i.e. 
		\begin{align}
			\Psi(\mbd{r}) =  e^{iq_x x + iq_y y} \begin{pmatrix}
			 u_{A\uparrow} \\[4pt]
			 u_{B\uparrow} \\[4pt]
			 v_{A\downarrow} \\[4pt]
			 v_{B\downarrow}
			\end{pmatrix}.
			\label{eq:wavefunc-realspace}
		\end{align}
Please note that the solutions of Eq. (\ref{eq:DBdG-eq}) in the SC region may be either an evanescent mode which decays exponentially with the increase in the distance from the interface at the energy $\epsilon$ of the incident electron smaller than the superconducting gap, namely $\epsilon < \Delta_\text{gap}$ or a propagating mode at $\epsilon > \Delta_\text{gap}$. Furthermore, since the Hamiltonian in the normal metal is diagonal in the spin subspace, here we only consider the incident electrons to be spin-up in Eq. (\ref{eq:wavefunc-realspace}).

Here, we denote $\psi(\mbd{q}) \equiv (u,~v) =  (u_{A\uparrow},~ u_{B\uparrow},~v_{A\downarrow},~ v_{B\downarrow})$. Therefore, the total wavefunction $\Psi_N \, (\Psi_{sc})$ in the N (SC) region can be expressed as a superposition of various eigenstates of Eq. (\ref{eq:DBdG-eq}) with positive excitation energy at the region $x < 0 \,\, (x>0)$ :
		\begin{widetext}
		\begin{align}
			&\Psi_{N}(\mbd{r}) = \psi^{(e)}_{N} (q_x, \, q_y) e^{iq_x x + iq_y y} + r_e \psi^{(e)}_{N} (-q_x, \, q_y) e^{-iq_x x + iq_y y} + r_h \psi^{(h)}_{N} (q^{\prime}_x, \, q_y) e^{iq^{\prime}_x x + iq_y y}, \nn[8pt]
			&\Psi_{sc}(\mbd{r}) = t_e \, \psi^{(e)}_{sc} (\bar{q}_x, \, q_y) e^{i\bar{q}_x x + iq_y y} + t_h \psi_{sc}^{(h)} (-\bar{q}^{ \prime}_x, \, q_y) e^{-i\bar{q}^{\prime}_x x + iq_y y} .
		\end{align}
		\end{widetext}
		with the incident state being normalized to unity. $r_{e, \, h}$ and $t_{e, \, h}$ , which depend on the energy $\epsilon$ and the wavevector $\mbd{q}$ of the incident state, represent the reflection and transmission coefficient for the electron branch (with subscript $e$) and hole branch (with subscript $h$). Due to the assumption of elastic scattering, $\psi^{(e), \, (h)}_{N, \, sc}$ are the eigenstates of Eq. (\ref{eq:DBdG-eq}) with the same excitation energy $\epsilon$. Please note that the transverse component of the wavevector $q_y$ is a conserved quantity during the scattering process due to the translational symmetry in $y$ while the longitudinal components for the electron and hole (electron-like and hole-like) states $q_{x},~q^{\prime}_x$ ($\bar{q}_x,~\bar{q}_x^{\prime}$) can be determined via their (quasiparticle) dispersion relations at a given $\epsilon$ and $q_{y}$.
 
Due to the presence of a $\delta$-gauge field as well as the restriction that the DBdG equations are of first order, the wavefunction continuity at the interface as being widely used in various literatures under the assumptions of ideal N/SC interfaces is not an appropriate boundary condition. Detailed derivations for the boundary condition for our theory are provided in the following: the DBdG equations for the $\mbd{K}_-$-valley in the presence of an effective $\delta$-gauge field as shown in Eq. (\ref{eq:vectorA}) are given by
%\begin{align}
	%		\left[ H_{DBdG} (\mbd{r}) + V (\mbd{r})\right] \, \psi (\mbd{r}) = \epsilon \,\psi (\mbd{r}) 
	%	\end{align}  
%\begin{align}
   % \frac{3t}{2} \, & \partial_{x} f_{B\uparrow}(x) -\frac{3 }{2} \Delta_0 \, \partial_{x} g_{B\downarrow}(x)  -\delta (x) \delta t f_{B\uparrow} (x)= E f_{A\uparrow}(x)  \nn[4pt]
    %-\frac{3t}{2} \, & \partial_{x}  f_{A\uparrow}(x) +3 \Delta_0  g_{A\downarrow}(x) -\delta t \delta(x) \, f_{A\uparrow}(x) = Ef_{B\uparrow}(x). \nn[4pt]
       % - \frac{3t}{2} \, & \partial_{x}  g_{B \downarrow}(x) + \delta t \delta (x) g_{B \downarrow} (x)= E v_{A\downarrow}(x) \nn[4pt]
        %\frac{3t}{2} \,  & \partial_{x}  g_{A\downarrow}(x) + \frac{3 \Delta}{2} \, \partial_{x} f_{A \uparrow} (x) + \delta t \delta (x) \, g_{A \downarrow}(x) = E g_{B\downarrow}(x).
%\end{align}
\begin{widetext}
\begin{align}
	\begin{pmatrix}
		0 & i\hat{q}_x + \hat{q}_y - \frac{\delta t}{t} \delta(x) -\frac{\mu}{t} & 0 & -\frac{3\Delta_0}{2t} (i\hat{q}_x -\hat{q}_y)	\\[3pt]
-i\hat{q}_x + \hat{q}_y - \frac{\delta t}{t} \delta(x) -\frac{\mu}{t} & 0 & \frac{3\Delta_0}{t} & 0  \\[3pt]
0 & \frac{3\Delta_0}{t} & 0 &  \frac{\mu}{t}  - i\hat{q}_x - \hat{q}_y + \frac{\delta t}{t} \delta(x) \\[3pt]
		\frac{3\Delta_0}{2t} (i\hat{q}_x + \hat{q}_y) & 0 &  \frac{\mu}{t}  + i\hat{q}_x - \hat{q}_y + \frac{\delta t}{t} \delta(x)  & 0 
		\end{pmatrix} \begin{pmatrix}
			f_{A\uparrow} \\[3pt]
			f_{B\uparrow} \\[3pt]
			g_{A\downarrow}  \\[3pt]
			g_{B\downarrow}
		\end{pmatrix} = 
		\frac{\epsilon}{t} \begin{pmatrix}
			f_{A\uparrow} \\[3pt]
			f_{B\uparrow} \\[3pt]
			g_{A\downarrow}  \\[3pt]
			g_{B\downarrow}
		\end{pmatrix}.
		\label{eq:DBdG-matrix}
\end{align}
\end{widetext}
Taking the momentum operator $\hat{q}_i \rightarrow -i\partial_i$ and integrating along the $x$-direction over a small distance across the interface, Eq. (\ref{eq:DBdG-matrix}) become
\begin{align}
    & \left[f_{B\uparrow}^{sc}(0) - f_{B\uparrow}^{N}(0) \right]-\frac{3 \Delta_0 }{2t} \, \left[ g_{B \downarrow}^{sc}(0) - g_{B \downarrow}^{N}(0)\right]  =  \frac{\delta t}{t} f_{B\uparrow} (0), \nn[4pt]
      & \left[f_{A\uparrow}^{N}(0) - f_{A\uparrow}^{sc}(0)\right]  = \frac{\delta t}{t}  \, f_{A\uparrow}(0).\, \nn[4pt]
         & \left[g_{B \downarrow}^{sc}(0)- g_{B \downarrow}^{N}(0) \right]  = \frac{\delta t}{t} \,  g_{B \downarrow} (0), \nn[4pt]
         & \left[g_{A \downarrow}^{N}(0)- g_{A \downarrow}^{sc}(0) \right]  + \frac{3 \Delta_0}{2t} \, \left[f_{A \uparrow}^{N}(0)- f_{A \uparrow}^{sc}(0) \right]   = \frac{\delta t}{t}  \, g_{A \downarrow}(0).
        \label{eq: BC-integrate}
\end{align}
In Eq. (\ref{eq: BC-integrate}), $f \, (g)_{\alpha \sigma} (0^{+}) \equiv f \, (g)_{\alpha \sigma} ^{sc}$ while $f \, (g)_{\alpha \sigma} (0^{-}) \equiv f \, (g)_{\alpha \sigma} ^{N}$ with $\alpha$ denotes the $A, \,\, B$ sublattice and $\sigma$ denotes the spin $\sigma = \uparrow, \, \downarrow$. 
%Furthermore, due to the translational invariance, the integration over $y$ vanishes. 
As we shall see in Eq. (\ref{eq: BC-integrate}), the boundary condition results in an ambiguity of $\Psi$ at $x=0$ which results from the situation that the DBdG equations are of first order. We further impose the following conditions to resolve the wavefunction ambiguity: in the situation of $\Delta_0 = 0$ and $\delta t = -t$ [\onlinecite{Castro2009strain}], we expect no tunneling current across the N/SC junction. The conditions are quite straightforward: in the case mentioned above, the original N/SC junction reduces to two disconnected semi-planes of graphene sheet; thus, electron tunneling is forbidden, giving rise to no tunneling current. The issue of wavefunction ambiguity at the origin now changes to problem of the electron tunneling across a junction of two pure graphene semi-planes with a $\delta$-gauge field taking the form of Eq. (\ref{eq:vectorA}) in between. Based on the condition of current conservation and the requirement of no tunneling current across the junction at $t = -\delta t$ and $\Delta_0 = 0$, we choose the wavefunctions at the origin as 
\begin{align}
	&f_{A\sigma} (0) = f_{A \sigma}^{sc},~~ f_{B\sigma} (0) = f_{B\sigma}^{N},\nn[3pt]
	& g_{A\sigma} (0) = g_{A\sigma}^{sc},~~g_{B \sigma} (0) = g_{B \sigma}^{N},
	\label{eq:BC-ambiguity}
\end{align}
where $\sigma = \uparrow,\downarrow$ denotes spin. Please note that due to the relation of time reversal partner between the incident electrons and the reflected holes within the DBdG formalism [\onlinecite{Beenakker2006a}, \onlinecite{Suzuura2002}]  for AR in a single-layered graphene, the hole wavefunctions $g$ share the same boundary conditions with the electrons wavefunctions $f$ as shown in Eq. (\ref{eq:BC-ambiguity}).  leading to
\begin{align}
   f_{B \uparrow}^{sc}(0) & = \eta f_{B \uparrow}^{N}(0)  +  \frac{ 3\Delta_0}{2t} \, \left[ g_{B \downarrow}^{sc}(0) - g_{B \downarrow}^{N}(0)\right], \nn[4pt]
     f_{B \downarrow}^{sc}(0) & = \eta f_{A \uparrow}^{sc}(0),   \nn[4pt]
         g_{B \downarrow}^{sc}(0) & = \eta g_{B \downarrow}^{N}(0), \nn[4pt]
         g_{A \downarrow}^{N}(0) & = \eta g_{A \downarrow}^{sc}(0)  +  \frac{ 3\Delta_0}{2t} \, \left[ f_{A \uparrow}^{sc}(0) - f_{A \uparrow}^{N}(0)\right],
        \label{eq: BC-final}
\end{align}
where $\eta \equiv 1 + \delta t/t$. For the detailed derivations for the boundary conditions in Eq. (\ref{eq:BC-ambiguity}), we refer the readers to the Appendix \ref{app:BC-graphene}. Once the reflection and transmission coefficients are obtained, following the BTK formalism [\onlinecite{BTK1982}], the normalized differential conductance can be computed by summing over all possible incident states, leading to 
\begin{align}
	\frac{G}{G_0} = \int^{\pi/2}_0 \, d \theta \, \cos \theta \left[	1- |r_e (eV, \, \theta )|^2 +| r_h (eV, \, \theta)|^2	\right],
\end{align}
where $G_0$ is the ballistic conductance of graphene [\onlinecite{Beenakker2006a}, \onlinecite{Jiang2008}].

\begin{figure}[ht]
	\includegraphics[scale=0.32]{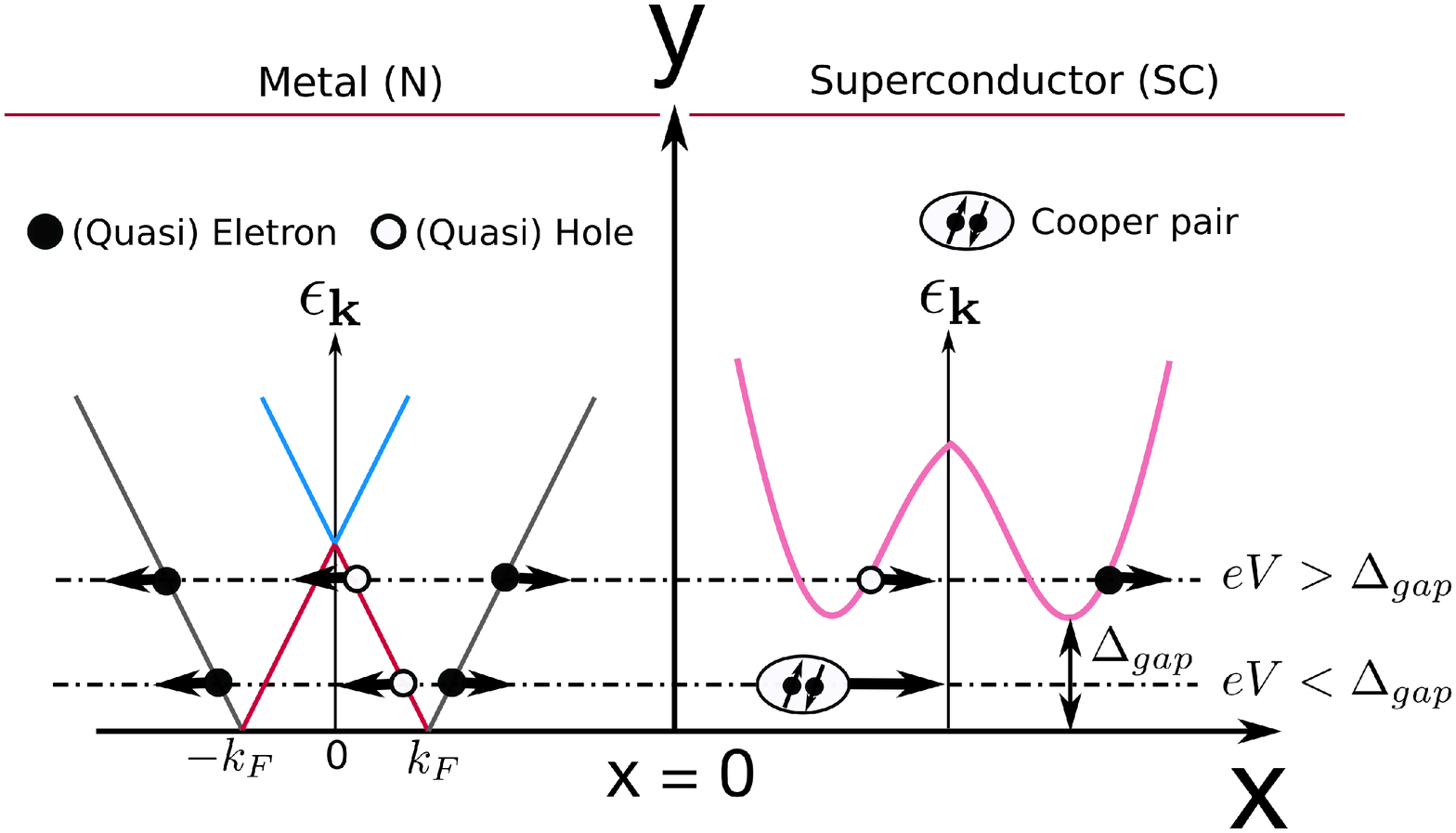}
	\caption{Schematically plot for the band structure in N and SC region at N/SC interface. In N region, the black solid line indicates the band structure for particles, while the red and blue solid line for the conduction- and valence-band holes. The Dirac-Bogoliubov quasiparticle dispersion is shown in SC region The direction of the arrows represents the direction of group velocity. This figure describes a general electron-hole conversion process at different bias $eV$.  }
	\label{fig:BTK-scattering}
\end{figure}

\begin{figure}[ht]
   \includegraphics[scale=0.45]{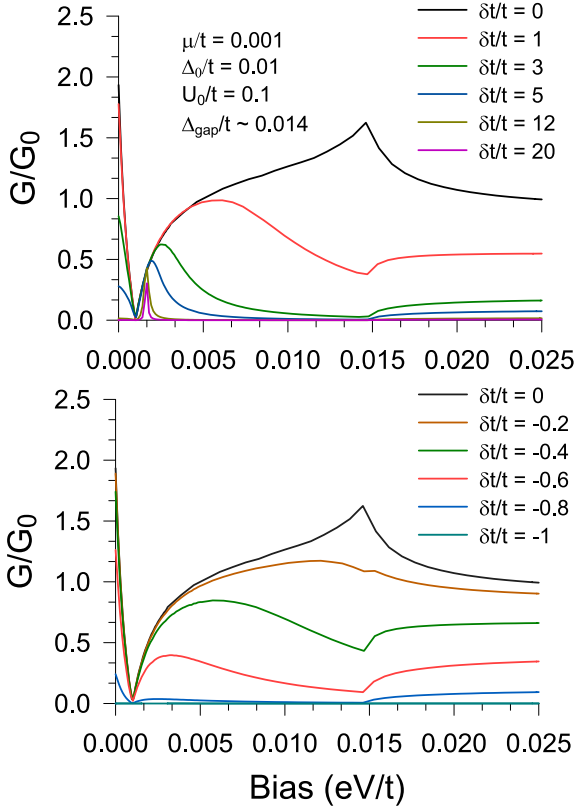}
   \caption{The Andreev conductance of a graphene - $d+id^{\prime}$ superconducting junction with varying barrier height $\delta t/t$ and fixed Fermi energy $\mu/t = 0.001$ and superconducting pairing strength $\Delta_{0}/t = 0.01,~\text{and}~U_{0}/t = 0.1$. The superconducting gap energy is around $\Delta_{gap}/t = 0.014$.}
   \label{fig-hu-dt-2}
\end{figure}

\begin{figure}[ht]
   \includegraphics[scale=0.45]{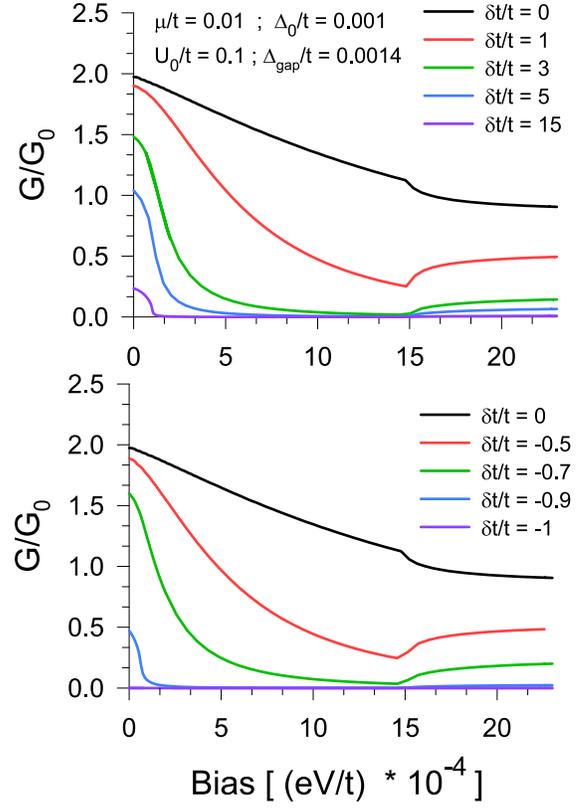}
   \caption{The normalized Andreev conductance $G/G_{0}$ with varying barrier height $\delta t/t$ and fixed Fermi level $\mu/t = 0.01$ and superconducting pairing strength $\Delta_{0}/t = 0.001$. The superconducting gap energy is found to be $\Delta_{gap}/t = 0.0014$.}
   \label{fig-hu-dt-1}
\end{figure}

In the absence of the local strain $\delta t = 0$, the Andreev conductance for the graphene $d+id^{'}$-wave superconductor N/SC junction is reproduced\cite{Jiang2008} as shown in Fig. \ref{fig-hu-dt-1} and \ref{fig-hu-dt-2} and the specular-AR to retro-AR transition marked by $eV/t = \mu$ can be easily identified for the case of $\Delta_{gap} > \mu$. The behaviors of the normalized differential conductance $G/G_{0}$ in Fig. \ref{fig-hu-dt-1} and \ref{fig-hu-dt-2} can be qualitatively explained via the aspect of the linear band structure of graphene normal metal and the electron-hole conversion processes at different Fermi energy as shown in Fig. \ref{fig:BTK-scattering}. For the case of $\mu < \Delta_{gap}$ as in Fig. \ref{fig-hu-dt-2}, at zero bias $eV = 0$, the phase spaces for the incident electrons and reflected holes are identical to each other, giving rise to the maximum Andreev conductance. However, once $eV$ is increased but still stays lower than $\mu$, it is clear that the phase space of hole band shrinks and results in a monotonic decline in $G/G_0$ until the bias reaches the Fermi energy $eV = \mu$, where there is no density of states for reflected hole, thus leading to a conductance dip as shown in Fig. \ref{fig-hu-dt-1}. In the regime $eV < \mu$, a conduction band electron is reflected as a conduction-band holes via the Andreev reflection, it is called the intraband Andreev reflection (or Andreev retro-reflection). Once the bias exceeds the Fermi energy $eV > \mu$, the incident electrons from the conduction band are converted as valence-band holes, leading to the interband Andreev reflection (or Andreev specular reflection).  For this case, $G/G_0$ increases again with the increasing $eV$ due to the increase of phase space in the hole band. At $eV \gg \Delta_{gap}$, the tunnelling process returns to the normal metal-normal metal tunnelling and the Andreev conductance saturates. Apparently, the Fermi energy serves as the "critical energy" for the transition between the intra to interband Andreev reflection. On the contrary, only the Andreev retro-reflection process exists as $\Delta _{gap} < \mu$, thus leading to a monotonic decline of $G/G_0$ as shown in Fig. \ref{fig-hu-dt-1}.

 In the following, we discuss the normalized conductance for the situation of non-zero barrier $\delta t \neq 0$. The vanishing of $G/G_0$ for the case of $\delta t = -t$ signifies no electron tunneling, as expected. For the situation of finite potential barrier, our results qualitatively capture the most significant features of Andreev conductance in the presence of $\delta$-barrier:  in Fig. (\ref{fig-hu-dt-1}) and (\ref{fig-hu-dt-2}), as the barrier strength $\delta t/t$ is increased $G/G_0$ dramatically decreases down to zero due to heavily scattering of electrons by the potential barrier, in well agreement with the previous results for the case of the $1d$ metal/$s$-wave N/SC junction in Ref. \onlinecite{BTK1982}. The cusps for the normalized conductance in Fig. \ref{fig-hu-dt-2} at $eV < \Delta_{gap}$ with non-zero $\delta t$ can be simply understood as the competition between the graphene density of states and the effect of the $\delta$-barrier: the density of states for graphene at low energy is linearly proportional to the excitation energy, signifying that increasing the energy will enhance the conductance. On the contrary, the effect of $\delta t$ tends to suppressed the conductance. Therefore, the competition between the density of states and the $\delta$-barrier gives rise to the cusps in Fig. \ref{fig-hu-dt-2}.
%. Furthermore, the Andreev conductance goes back to the results obtained in  as the potential is switched off.

%%%%%%%%%%%%%%%%%%%%%%%%%%%%%%%%
%%%%%%%%%%%%%%%%%%%%%%%%%%%%%%%%
%%%%%%%%%%%%%%%%%%%%%%%%%%%%%%%%
\section{AR across a Kane-Mele/$d+id^{\prime}$-wave N/SC junction }
In the past decades, much effort has been devoted to searching for the novel topological states of matters. The two examples of particular interest are the topological insulators [\onlinecite{Qi2011RMP}, \onlinecite{Hasan2010RMP}], which have insulating bulk states while the edge or surface supports time-reversal symmetry protected conducting states, and the topological superconductors which support gapless, charged neutral Majorana edge (or surface) states [\onlinecite{Alicea-MF}] with superconducting bulk states. 

Recently, the doped Kane-Mele (KM) model, which was originally proposed in Ref. \onlinecite{Kane2005PRL-1} and \onlinecite{Kane2005PRL-2}, with large onsite electron-electron repulsive interaction on a $2d$ periodic honeycomb lattice has been theoretically shown to feature time-reversal broken $d_{x^{2}-y^{2}}+id_{xy}$-wave superconducting state in the bulk via renormalized mean-field theory [\onlinecite{Schaffer2007PRB}, \onlinecite{SchafferRev-chiral-dwave}]. Moreover, it was also found that the system undergoes a topological phase transition from the helical superconducting to the chiral superconducting order as the strength of the intrinsic spin-orbit coupling is decreased, and two pairs of counter-propagating helical Majorana zero modes have been found theoretically at the edges of a finit-sized zigzag ribbon of the tight-binding KM $t$-$J$ model in spite of the time-reversal broken $d_{x^{2}-y^{2}}+id_{xy}$-wave superconducting order [\onlinecite{Yung2015}, \onlinecite{Mou-duality}]. Via the numerical simulation by density functional theory, the KMtJ model may be realized via doping adatoms such as indium or thallium on a graphene sheet, which generates an effective Kane-Mele type intrinsic SO coupling ($\sim 20$ meV) [\onlinecite{Alicea-Adatom}], more larger than the un-doped graphene. Besides graphene-based systems, our model is also applicable to other compounds with underlying honeycomb lattice such as In$_3$Cu$_2$VO$_9$ [\onlinecite{LeHur-dwave-honeycomb}, \onlinecite{Moller-In-compound, Yan-In-compound, Liu-In-compound}], $\beta$-Cu$_2$V$_2$O$_7$[\onlinecite{Tsirlin-Cu-compound}, \onlinecite{KTLaw-Cu-compound}], MoS$_2$ [\onlinecite{Ye-SC-dome}] and silicene [\onlinecite{Silicene-d-wave}]. Those materials have been proposed to exhibit chiral-$d$-wave superconducting state around half-filling. These exotic features discovered in the KMtJ model motivates us to seek the corresponding experimental signatures. 

In the following, we investigate the Andreev reflection across a N/SC junction with the normal side being modeled by the doped KM model while the SC region is a doped correlated KM $t$-$­J$ model with $d+id^{\prime}$-­wave spin-singlet superconducting order. 
%%%%%%%%%%%%%%%%%%%%%%%%%%%%%%%%%%%%%%
%%%%%%%%%%%%%%%%%%%%%%%%%%%%%%%%%%%%%%
%\subsection{The KM $tJ$ N/SC junction}
The Kane-Mele model which can be viewed as a spinful Haldane model [\onlinecite{Haldane1988}] is composed of the nearest-neighbor (NN) tight-binding Hamiltonian $H_0$ as in Eq. (\ref{eq:GrapheneTB}) and the next-nearest neighbor (NNN) hopping intrinsic spin-orbit (SO) interaction $H_{SO}$:  
\begin{align}
    &H_{KM} = H_O + H_{SO} + H_{\mu}, \nn
    &H_{SO} = \, i \lambda_{SO} \sum_{\ll \, i, \, j \gg} \sum_{\sigma, \, \sigma^{\prime} = \uparrow \downarrow}  \, \nu_{ij} \sigma^{z}_{\alpha \alpha^{\prime}}  \, c_{i \alpha}^{\dagger}c_{j \alpha^{\prime}}, \nn
   &H_\mu  = -\mu \sum_{i,\sigma} c_{i\sigma}^\dagger c_{i\sigma}.
   \label{eq:KM}
 \end{align}
Here, $c_{i\sigma} (c_{i\sigma}^{\dagger})$ annihilates (creates) an electron on either the $A$ or $B$ sublattice on the $i$-th unit cell. $\ll i,j \gg$ denotes the NNN indices, $\lambda_{SO}$ is the coupling strength of the intrinsic SO interaction and $\sigma,\, \sigma^\prime = \uparrow,\,\downarrow$ represent spins. $\nu_{ij} = \pm 1$ is an orientation dependent factor: $\nu_{ij} = 1 $ for an electron makes a right turn while moves from the $i$-th site to its $j$-th NNN site, and $\nu_{ij} = -1 $ for left turn. The doping is characterized by $H_\mu$ with $\mu$ being the value of the chemical potential. The mean-field Hamiltonian on a periodic lattice in terms of the basis $\Psi_{\mbd{k}} = \left( c_{A,\mbd{k}}^{\uparrow} \, c_{B,\mbd{k}}^{\uparrow}, \, c_{A,\mbd{k}}^{\downarrow} \, c_{B,\mbd{k}}^{\downarrow}, \, c_{A,-\mbd{k}}^{\uparrow \dagger} \, c_{B,-\mbd{k}}^{\uparrow \dagger}, \, c_{A,-\mbd{k}}^{\downarrow \dagger} \, c_{B, -\mbd{k}}^{\downarrow \dagger}\right)^T$ is given by the $ 8\times 8$ matrix: 
\begin{align}
    \mathcal{H}_{\mbd{k}} =& \, \left( \begin{matrix}
         h^{+}_{\mbd{k}} - \mu  &   0  & 0 & \Delta_{\mbd{k}}  \\[3pt]
         0 & h^{-}_{\mbd{k}} -  \mu & -\Delta_{\mbd{k}} & 0 \\[3pt]
         0 &  -\Delta_{\mbd{k}}^{\dagger} &  \mu-h_{-\mbd{k}}^{+^*}  & 0\\[3pt]
         \Delta_{\mbd{k}}^{\dagger}  & 0 & 0 & \mu -h_{\mbd{-k}}^{-^*}
    \end{matrix}\right) \nn
     =& \, \left( \begin{matrix}
         h_{\mbd{k}}^{+} - \mu  &   0  & 0 & \Delta_{\mbd{k}}  \\[3pt]
         0 & h_{\mbd{k}}^{-} -  \mu & -\Delta_{\mbd{k}} & 0 \\[3pt]
         0 &  -\Delta_{\mbd{k}}^{\dagger} &  \mu-h_{\mbd{k}}^{-}  & 0\\[3pt]
         \Delta_{\mbd{k}}^{\dagger}  & 0 & 0 & \mu -h_{\mbd{k}}^{+}
    \end{matrix}\right) 
    \label{eq:KMtJ-H}
\end{align}
with
\begin{align}
    &h_{\mbd{k}}^{\pm} = \begin{pmatrix}
    	\pm \gamma(\mbd{k}) & f (\mbd{k}) \\[3pt]
    	f^{\ast}({\mbd{k}}) & \mp\gamma({\mbd{k}}) 
\end{pmatrix} ,  \\[3pt]
& \gamma({\mbd{k}}) =  2\lambda_{SO} \left[ 2 \,  \text{cos}\frac{3k_{x}}{2}  \text{sin}\frac{\sqrt{3} k_{y}}{2}- \text{sin}\sqrt{3} k_{y}\right]. 
\end{align}
In the second line of Eq. (\ref{eq:KMtJ-H}), we have applied the relations of $\gamma({-\mbd{k}} )= -\gamma(\mbd{k})$ [\onlinecite{Mott-TI-LeHur}] and $f^\ast({\mbd{k}})= f({-\mbd{k}}) $. 

The electronic excitations near the $\mbd{K}_\tau$-valley in the N and SC sides are described by the linearized DBdG equations, which take the form
\begin{align}
			 \begin{pmatrix}
          \mathcal{H}_{\tau}(\mbd{q} )  - \mu & \Theta(x) \bar{\Delta}_{\tau}(\mbd{q} ) \\[4pt]
         \Theta(x)  \bar{\Delta}_{\tau}(\mbd{q} )^{\dagger} & \mu  - \mathcal{H}_{\tau}(\mbd{q} )
    \end{pmatrix}  \begin{pmatrix}
    	u_\tau \\[4pt]
    	v_\tau
    \end{pmatrix} = \frac{\epsilon_{\mbd{q}}}{t}   \begin{pmatrix}
    	u_\tau \\[4pt]
    	v_\tau
    \end{pmatrix}.
    \label{eq:DBdG-eq-KMtJ}
		\end{align}
Here, $(u_\tau,~v_\tau) = (u_{A\tau}^\uparrow, u_{B\tau}^\uparrow, u_{A\tau}^\downarrow, u_{B\tau}^\downarrow, v_{A\tau}^\uparrow, v_{B\tau}^\uparrow, v_{A\tau}^\downarrow, v_{B\tau}^\downarrow  )^T$ is an eight-component wavefunctions in the momentum domain near the $\mbd{K}_\tau$-valley with the first four components $u_\tau$ for particles while the last four components $v_\tau$ for holes.
\begin{align}
	  \mathcal{H}_{\tau}(\mbd{q} ) =&\, -\sigma_0 ( \pi^{x} q_{y} + \tau \pi^{y} q_{x}) - 3\sqrt{3}\,   \lambda (x) \, \tau \sigma^z \pi^z\nn
	  & -(U_0 /t) \sigma^0 \pi^0 \Theta(x),
\end{align}
where the $2\times 2 $ unit matrix $ \sigma^0 = \text{diag}(1,~1)$ together with the three Pauli matrices $\sigma^{x,y,z}$ are for the spin subspace in the Hilbert space while the matrices $\pi$ are for the sublattices as already defined in Eq. (\ref{eq:Pauil-sublattice}) and Eq. (\ref{eq:tau0-sublattice}).  Here, we assume the magnitudes of intrinsic SO coupling in the N and the SC region can be adjusted independently, thus we introduce
\begin{align}
\lambda (x) = 
	\begin{cases}
		\lambda_{SO}/t , ~~\text{for}~x<0, \\[4pt]
		\lambda^{\prime}_{SO}/t, ~~\text{for}~x>0,
	\end{cases}
\end{align}
where $\lambda_{SO}$ indicates the intrinsic SO coupling in N while ${\lambda}^{\prime}_{SO}$ in SC. Here, $\bar{\Delta}_\tau (\mbd{q})$ is a $4 \times 4$ matrix for the linearized $d+id^{\prime}$-wave pairing near the $\mbd{K}_\tau$ Dirac point, which takes the form
	\begin{align}
	&\bar{\Delta}_\tau (\mbd{q}) \equiv \begin{pmatrix}
		    0 & \Delta_\tau (\mbd{q})  \\[4pt]
		    -\Delta_\tau (\mbd{q})  & 0
	       \end{pmatrix}.
		\end{align}
Around $\mbd{K}_-$, $\Delta_- (\mbd{q}) $ is given by the Eq. (\ref{eq:d+id-linearized}) in the previous section, which is related to the one for the $\mbd{K}_+$-valley by $\Delta_+ (\mbd{q}) = \Delta_{-}^{T} (-\mbd{q})$.

\begin{figure*}[ht]
   \includegraphics[scale=0.6]{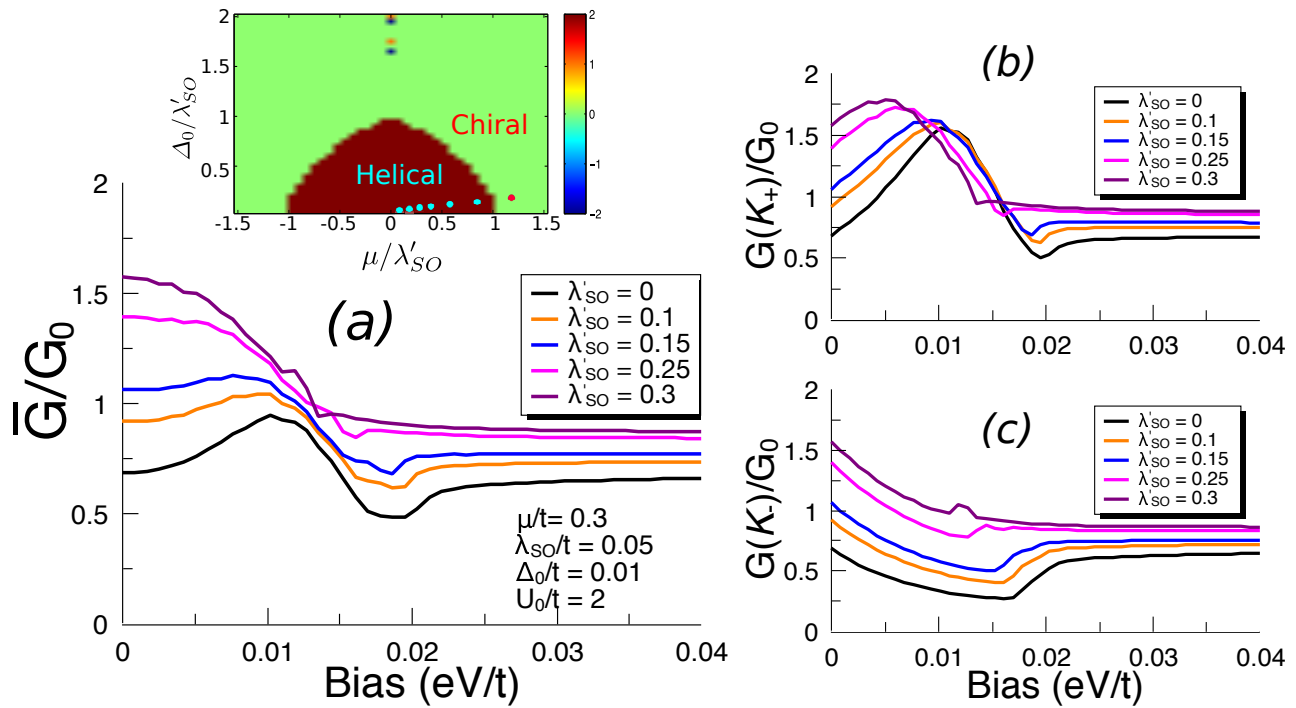}
   \caption{The plot (a) shows the average normalized Andreev conductance $\bar{G}/G_0$ across a KM - $d+id^{\prime}$-wave N/SC junction in terms of varying  the intrinsic SO coupling $\lambda^{\prime}_{SO}$ in the SC side in the absence of local strain. Here, we fix the Fermi energy $\mu/t = 0.3$, the intrinsic SO coupling $\lambda_{SO}/t = 0.05$ in N region, the electrostatic potential  $U_0/t = 2$ in SC and the value of SC pairing $\Delta_{0}/t = 0.01$. The inset illustrates the topological phase diagram for the bulk state in SC region, showing that the bulk will undergo the chiral (green area)-to-helical (brown area) topological transition as the ratio $\Delta_0/\lambda_{SO}^{\prime}$ or $\mu/\lambda_{SO}^{\prime}$ is varied. The color bar represents the value of the spin-Chern number. Plots (b) and (c) show the normalized Andreev conductance $G(K_+)/G_0$ and $G(K_-)/G_0$ contributes from the electrons from the $\mbd{K}_+$ and $\mbd{K}_-$-valley, respectively.}
   \label{fig:KMtJ-mixed-nostrain}
\end{figure*}

\begin{figure*}[ht]
   \includegraphics[scale=0.6]{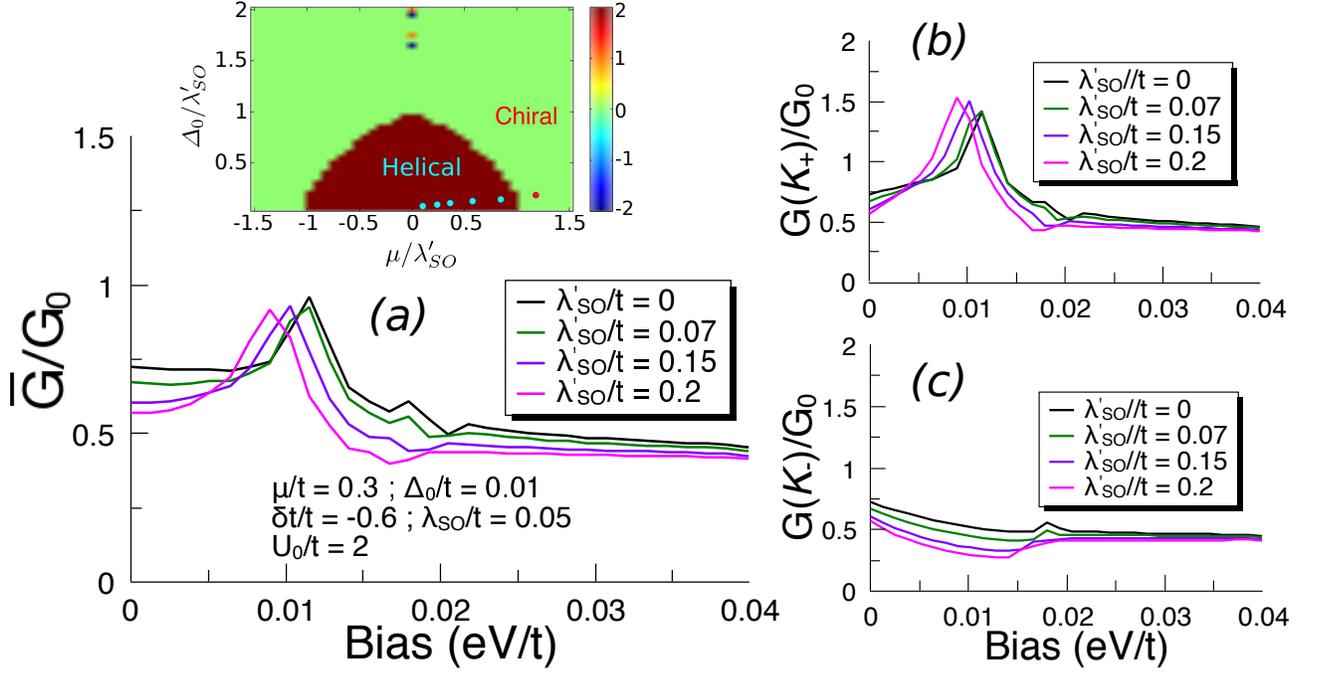}
   \caption{Figure (a) shows the average Andreev conductance $\bar{G}/G_0$ in the presence of a constant $\delta$-gauge field with barrier strength $\delta t/t = -0.6$ at the interface, while (b) and (c) for the Andreev conductance $G(K_\pm)/G_0$ from the $\mbd{K}_{\pm}$-valley, respectively.  All the remaining parameters are the same as that in Fig. \ref{fig:KMtJ-mixed-nostrain}.}
   \label{fig:KMtJ-mixed-strain}
\end{figure*}

For the reason that the KM$tJ$ model exhibites an effective spin-singlet $p \pm i p^{\prime}$ superconducting order near the two Dirac points $\mbd{K}_\pm$ [\onlinecite{Mou-duality}], the valley degeneracy no longer exist and we ought to consider the normalized Andreev conductance contributed from $\mbd{K}_+$ and $\mbd{K}_-$, respectively. Here, we assume the electron scattering only occurs within one valley, thus the normalized Andreev conductance can be simply evaluated by taking the average of the individual contributions from $\mbd{K}_\pm$. The Andreev conductance contributed from one valley can be similarly obtained via the BTK formalism. The average of the normalized Andreev conductance is expressed as 
\begin{align}
	\frac{\bar{G}}{G_0} = \frac{1}{G_0} \cdot \frac{G(K_+) + G(K_-)}{2},
\end{align}
where $G(K_{\pm})$ is the Andreev conductance from the $\mbd{K}_\pm$, respectively. The results are illustrated in Fig. \ref{fig:KMtJ-mixed-nostrain} and Fig. \ref{fig:KMtJ-mixed-strain}.
\begin{figure}
	\centering
	\includegraphics[scale=0.35]{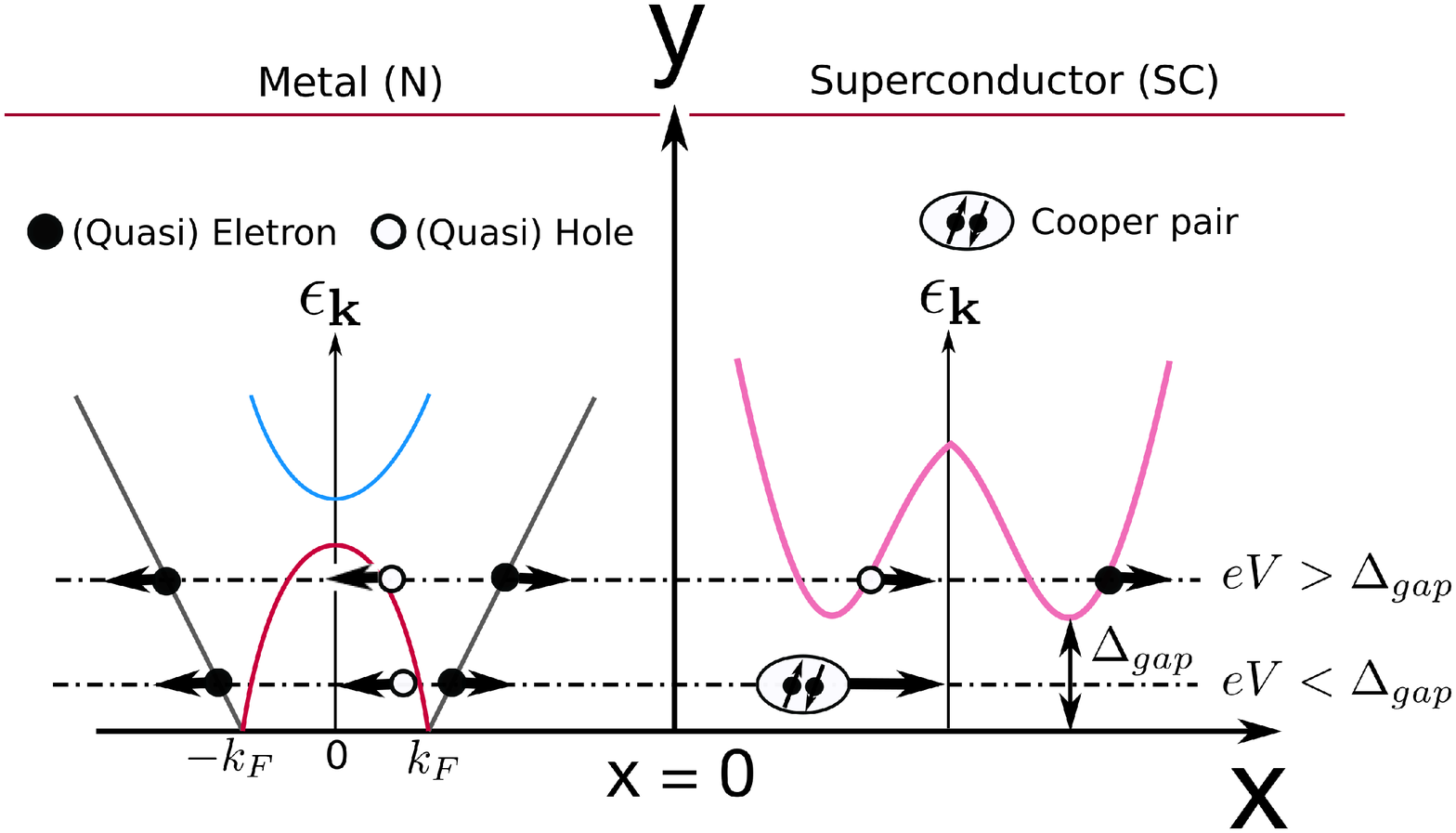}
	\caption{The band structure in the left region is for the Kane-Mele model near the Dirac points. Due to the intrinsic SO coupling, we can see a band gap between the conduction (red solid line) and valence (blue solid line) hole bands.}
	\label{fig:BTKscattering-2}
\end{figure}

In the following, we qualitatively discuss the conductance behaviors for Fig. \ref{fig:KMtJ-mixed-nostrain} and \ref{fig:KMtJ-mixed-strain}. As the $\delta$-barrier is switched off, the Andreev conductance $G(K_-)$ from the $\mbd{K}_-$-valley, as shown in Fig. \ref{fig:KMtJ-mixed-nostrain}-(c), monotonically decreases with the increasing bias $eV$ due to the shrinking of the phase space in the hole band as shown in Fig. \ref{fig:BTKscattering-2}.
% since the Fermi energy in this case  is higher than the SC gap energy $\mu > \Delta_{gap}$.
  The resulting behavior of the Andreev conductance $G(K_-)$ is similar to the case in Fig. \ref{fig-hu-dt-1}, hence only the Andreev retro-reflection process involves. Due to the inversion symmetry breaking in the KMtJ model, the Andreev conductance $G(K_+)$ from the $\mbd{K_+}$ in Fig. \ref{fig:KMtJ-mixed-nostrain}-(b) behaves entirely different from $G(K_-)$.
  % In our case with $\mu/t = 0.3 \gg \Delta_0/t$, 
  Averaging $G(K_+)$ and $G(K_-)$ gives rise to distinctive Andreev conductance behaviors for different values of $\lambda_{SO}^\prime$.
   %The chiral to helical topological phase transition takes place around $\lambda_{SO}/t^\prime \sim 0.1$ (corresponds to the orange line in Fig. \ref{fig:KMtJ-mixed-nostrain}). 
   For small intrinsic SO coupling,  we find that the average Andreev conductance $\bar{G}/G_0$ increases as the bias is increased at low bias, which $\bar{G}/G_0$ behaves in a similar manner to the Andreev specular reflection. On the contrary, for large intrinsic SO coupling $\lambda_{SO}^\prime$, $\bar{G}/G_0$ monotonically decreases with increasing bias. We argue the Andreev retro-reflection may dominate $\bar{G}/G_0$ in this situation.

%\begin{figure}[ht]
   %\includegraphics[scale=0.5]{fig-BTK-KM-kmtj-tso-00015}
   %\caption{The Andreev conductance with varying barrier height $dt/t$ and fixed Fermi energy $E_{F}/t = 0.01$, $t_{SO}/t = 0.0015$ and superconducting pairing strength $\Delta_{0}/t = 0.001$}
%\end{figure}
%\begin{figure}[ht]
   %\includegraphics[scale=0.5]{fig-BTK-KM-kmtj-dt-m06}
 %  \caption{The Andreev conductance with varying barrier height $dt/t$ and fixed Fermi energy $E_{F}/t = 0.01$, $t_{SO}/t = 0.0015$ and superconducting pairing strength $\Delta_{0}/t = 0.001$}
%\end{figure}
%\begin{figure}[ht]
   %\includegraphics[scale=0.7]{fig-BTK-KM-kmtj-Ef-0001-Delta-001-tso-00001}
   %\caption{The Andreev conductance with varying barrier height $dt/t$ and fixed Fermi energy $E_{F}/t = 0.01$, $t_{SO}/t = 0.0015$ and superconducting pairing strength $\Delta_{0}/t = 0.001$}
%\end{figure}

%%%%%%%%%%%%%%%%%%%%%%%%%%%%%%%%%
%%%%%%%%%%%%%%%%%%%%%%%%%%%%%%%%%

\section{Discussion and Conclusion}
Before we conclude, the effect of electron scattering by the edge states on the Andreev conductance deserves some discussions here. As mentioned in the previous section,  via the bulk-edge correspondence [\onlinecite{Hasan2010RMP}, \onlinecite{Moore-TI}], the KMtJ model supports chiral or helical Majorana edge states depending on the topological phase in the bulk. Accordingly, we anticipate that as the electron scattering by the quasiparticles in the edge states at the N/SC interface is considered, the Andreev conductance may exhibit distinctive behavior rather than a relatively smooth crossover as the SC region is tuned to undergo chiral-to-helical topological phase transition. 
%The scattering of the incident electrons by different types of edge states leads to salient features in the Andreev conductance.
 We expect this distinctive behavior on the Andreev conductance at the topological critical point will serve as an experimentally signature to probe the topological phase transition.
 
In conclusion, we have investigated the Andreev reflection based on Blonder-Tinkham- Klapwijk formalism in a graphene normal metal/$d+id^\prime$-wave superconducting junction in the case of a finite barrier laying on the N/SC interface. In order to investigate the electron scatterings on a N/SC junction with different transparencies, an effective Dirac $\delta$-guage field has been introduced by applying a homogeneous local strain parallel to the interface. In the absence of local strain, i.e $\delta t = 0$, our results successfully reproduce the normalized Andreev conductance $G/G_0$ curves in Ref. \onlinecite{Jiang2008}. At the other extreme parameter regime of $\delta t = -t$, the Andreev conductance vanishes because the N/SC junction is disconnected and therefore electron tunneling is forbidden. For the case of finite barrier and at the energy of the incident electron lower than the superconducting gap $\Delta_{gap}$, the Andreev conductance dramatically decreases down to zero as the barrier strength $\delta t$ is increased since the incident electrons scatter heavily with the barrier. Our results are qualitatively consistent with the results of the Andreev conductance originally obtained for the case of a $1d$ normal metal/$s$-wave superconducting junction.
% Furthermore, we apply the same formalism to a 2D quantum spin-Hall insulator on honeycomb lattice. 

We further investigate the Andreev reflection across the N/SC junction with the N region being described by the doped Kane-Mele model while the SC region features $d+id^{\prime}$-wave spin-singlet pairing induced by strongly electron correlations. The normalized Andreev conductance contributed from the $\mbd{K}_+$ and $\mbd{K}_-$-valleys illustrate entirely different behaviors due to the different effective superconducting pairing symmetry near the two Dirac points. Our results provide spectra of the normalized Andreev conductance within certain parameter regimes for future experiments.  

\section{Acknowledgement}
This work was supported by Ministry of Science and
Technology (MoST), Taiwan. We also acknowledge support
from TCECM and Academia Sinica Research Program
on Nanoscience and Nanotechnology, Taiwan. This work is also supported by the MOST grant No. 104-2112-M-009 -004 -MY3, the MOE-ATU program, the NCTS of Taiwan, R.O.C. (CHC).

\appendix
\section{Boundary conditions for the graphene N/N junction with a $\delta$-gauge field} 
\label{app:BC-graphene}
In this section, we repeat the derivations on the boundary conditions for the electronic transport through a graphene normal metal-normal metal (N/N) junction in the low-energy limit in the presence of a effective $\delta$-gauge field in between. This issue was originally addressed by Castro Neto $et.~al.$ in Ref. \onlinecite{CastroNeto2009RMP} and \onlinecite{Castro2009strain}.
\subsection{For the $\mbd{K}_-$-valley}
%The linearized Hamiltonian at the $\mbd{K}_{-}$-valley is given by
%\begin{align}
%-\int d^{2}r \, \Psi^{\dagger}_{-} (\mbd{r}) \left( \sigma_{y} \left( \hat{q}_{x} + \mathcal{A}_{x} \right) - \sigma_{x} \left( \hat{q}_{y}  + \mathcal{A}_{y}	 \right) \right)  \Psi_{-} (\mbd{r}) ,
%\end{align}
The Schrodinger's equations for the graphene tight-binding model in the low-energy limit near the $\mbd{K}_{-}$-valley read
\begin{align}
	&-\left[ \pi^y \left( -i\partial_x \right) -\pi^x \left(-i\partial_y +  \mathcal{A}_y\right) \, \right] \Psi_-(\mbd{r}) = (\epsilon/t) \, \Psi_-(\mbd{r}) \nn
	\hfill \nn
	\Rightarrow \, & \left[ \begin{matrix}
	 0 & -\partial_x + i \partial_y - \mathcal{A}_y \\
	\partial_x + i \partial_y - \mathcal{A}_y & 0
	\end{matrix}\right]  \Psi_-(\mbd{r}) = -(\epsilon/t) \, \Psi_-(\mbd{r}) .
	\label{eq:SE-Km}
\end{align}
Expressing the two-component wavefunction as $\Psi_{-}(\mbd{r}) = (\psi_{A}^{\prime} (\mbd{r}), ~\psi_{B}^{\prime} (\mbd{r}))^{T}$ and substituting the effective $\delta$-potential to $\mathcal{A}_y$ in Eq. (\ref{eq:SE-Km}) yields
\begin{align}
	\Rightarrow 
	\begin{cases}
		\left( -\partial_x + i \partial_y + \frac{\delta	t}{t} \delta(x) \right)  \psi_B^{\prime}(\mbd{r}) = -(\epsilon/t) \, \psi^{\prime}_A(\mbd{r}) \\
		\left( \partial_x + i \partial_y + \frac{\delta	t}{t} \delta(x) \right)  \psi_A^{\prime}(\mbd{r}) = -(\epsilon/t) \, \psi^{\prime}_B(\mbd{r}).
	\end{cases}
\end{align}
Integrating the Schrodinger's equations in Eq. (\ref{eq:SE-Km}) over a infinitesimal region across the origin, i.e. $\int^{0^+}_{0^-} \, dx$, we obtain the boundary condition 
\begin{align}
	\Rightarrow 
	\begin{cases}
	\psi^{\prime}_B(0^-) + \frac{\delta t}{t} \, \psi^{\prime}_B(0) = \psi^{\prime}_B(0^+), \\
	\psi^{\prime}_A(0^+) + \frac{\delta t}{t} \, \psi^{\prime}_A(0) = \psi^{\prime}_A(0^-).
	\label{eq:SE-BC-Km}
	\end{cases}
\end{align}
Since the Schrodinger's equations with linear Dirac spectrum are of first order, we cannot demand the wavefunction to be continuous at the origin, therefore giving rise to the wavefunction ambiguity, $\psi_{A}^{\prime} (0) $ and $\psi_{B}^{\prime} (0) $, as we shall see in Eq. (\ref{eq:SE-BC-Km}). Here, we choose the undetermined wavefunctions in following the way : 
\begin{align}
\psi^{\prime}_B(0) = \psi^{\prime}_B(0^-)~;~\psi^{\prime}_A(0 ) = \psi^{\prime}_A(0^+)
\label{eq:wavefunction-BC-Km}
\end{align}
and Eq. (\ref{eq:SE-BC-Km}) become
\begin{align}
	\Rightarrow
	\begin{cases}
		\psi^{\prime}_B(0^+) = \eta \, \psi^{\prime}_B(0^-), \\
		\psi^{\prime}_A(0^-) = \eta \, \psi^{\prime}_A(0^+).
	\end{cases}
	\label{eq:BCKminus-final}
\end{align}
Later, we will show that the choice of the undetermined wavefunctions at $x= 0$ in Eq. (\ref{eq:wavefunction-BC-Km}) will lead to the conservation of probability current. Thus, current conservation justify our choice of the undetermined wavefunctions. 

We may image that an incident electron far from the interface in the graphene sheet at the region of $x<0$ moves toward the interface and get scattered with the potential at the origin. By solving the Schrodinger's equation, the right-moving state in $\mbd{q}$-space for the incident electron is given by
\begin{align}
	\tilde{\Psi}_- (q_x,\,q_y) = \frac{1}{\sqrt{2}} \left( \begin{matrix}
		-\frac{iq_x + q_y}{q} \\
		-1
	\end{matrix}\right) = -\frac{e^{i\phi}}{\sqrt{2}} \left( \begin{matrix}
	1 \\
	e^{-i\phi}
\end{matrix}	
\right)
\label{eq:Km-incidentstate}
\end{align}
while the left-moving state for the reflected electron can be obtained by simply reverse the sign of $q_{x}$ in Eq. (\ref{eq:Km-incidentstate}), which is given by
\begin{align}
\tilde{\Psi}_-(-q_x,\,q_y) = \frac{1}{\sqrt{2}} \left( \begin{matrix}
		\frac{iq_x-q_y}{q} \\
		-1
	\end{matrix}\right) = -\frac{e^{-i\phi}}{\sqrt{2}} \left( \begin{matrix}
		1 \\
		e^{i\phi}
	\end{matrix}\right).
	\label{eq:Km-reflectedstate}
\end{align}
The total wavefunctions with normalized incident state in the left and right side of the graphene sheet are
\begin{widetext}
\begin{align}
		\begin{cases}
		\Psi^{L}_{-}(\mbd{r}) = e^{iq_x x+iq_y y } \, \tilde{\Psi}_-(q_x , q_y) + \mathcal{R}\, e^{-iq_{x}  x+iq_y y } \, \tilde{\Psi}_- (-q_x , q_y), \\
		\Psi^{R}_{-}(\mbd{r}) = \mathcal{T}\,e^{iq_{x} x+iq_y y } \,\tilde{\Psi}_-(q_x , q_y)
		\end{cases}
\end{align}
\end{widetext}
with $\mathcal{R}$ and $\mathcal{T}$ being denoted as the reflection and transmission coefficients. Here, the superscripts $L$ and $R$  stand for left and right, respectively. The boundary conditions are give by
\begin{align}
	\begin{cases}
		\mathcal{T} e^{-i\phi} = \eta \left( e^{-i\phi} +R e^{i\phi}	\right) ,\\
		1+\mathcal{R} = \eta \,\mathcal{T}.
	\end{cases}
\end{align}

\begin{align}
	\mathcal{T}= \eta \, \frac{1-e^{2i\phi}}{1-\eta^2 \, e^{2i\phi}}\quad ; \quad \mathcal{R}=\frac{\eta^2-1}{1-\eta^2 \, e^{2i\phi}}.
\end{align}
Apparently, $\mathcal{T}= 0$ when $\eta = 0$ as expected and the probability current is conserved, that is $|\mathcal{R}|^2 + |\mathcal{T}|^2 = 1$.

\subsection{For the $\mbd{K}_+$-valley}
Likewise, the Schrodinger equations for the $\mbd{K_{+}} = (0, ~-\frac{4\pi}{3\sqrt{3}})$-valley are given by
\begin{align}
	& \left[ \pi^{y \, *} \left( -i\partial_x \right) -\pi^x \left(-i\partial_y - \mathcal{A}_y \right) \, \right] \Psi_+(\textbf{r}) = (\epsilon/t) \, \Psi_+(\textbf{r}) \nn
	\hfill \nn
	\Rightarrow \, & \left[ \begin{matrix}
	 0 & \partial_x + i \partial_y +\mathcal{A}_y \\
	- \partial_x + i \partial_y +\mathcal{A}_y & 0
	\end{matrix}\right]  \Psi_+(\textbf{r}) = (\epsilon/t)\, \Psi_+(\textbf{r}) .
\end{align}
Express $\Psi_+(\textbf{r}) =\left( \psi_A (\textbf{r}), ~ \psi_B(\textbf{r}) \right)^T$, we have
\begin{align}
\Rightarrow
	\begin{cases}
		\left( \partial_x + i \partial_y + \mathcal{A}_y	\right) \psi_B(\textbf{r}) =(\epsilon/t) \, \psi_A (\textbf{r}), \\
		\left( -\partial_x + i \partial_y +\mathcal{A}_y	\right) \psi_A(\textbf{r}) =  (\epsilon/t) \, \psi_B (\textbf{r}).
	\end{cases}
\end{align}
Next, integrating from $x= 0^-$ to $x= 0^+$ yields the boundary conditions: 
\begin{align}
	\Rightarrow 
	\begin{cases}
		\psi_B(0^+) - \psi_B(0^-) = \frac{\delta t}{t} \, \psi_B(0), \\
		\psi_A(0^-) - \psi_A(0^+) =  \frac{\delta t}{t} \, \psi_A(0). 
		\label{eq:BC-Kp}
	\end{cases}
\end{align}
Apparently, we can immediately see the undetermined wavefunctions $\psi_A(0)$ and $\psi_B(0)$ appears on the R.H.S. in Eq. (\ref{eq:BC-Kp}). For the same reason of current conservation, we choose the undetermined wavefunction to be
\begin{align}
	\psi_B (0) = \psi_B(0^-);~\psi_A (0) = \psi_A(0^+).
	\label{eq:BC-ambiguity-Kp}
\end{align}
Please note that due to the relation of the time-reversal partner for the states near the $\mbd{K}_+$ and $\mbd{K}_-$ wavevector, the choice of the undertermined wavefunctions at $x=0$ in Eq. (\ref{eq:BC-ambiguity-Kp}) is identical to that in Eq. (\ref{eq:wavefunction-BC-Km}).

 To calculate the transmission and reflection coefficients, we first prepare the normalized right-moving and left-moving states for the incident and reflected electrons in the momentum space:
\begin{align}
	&\tilde{\Psi}_+(q_x,\,q_y) = \frac{1}{\sqrt{2}} \left( \begin{matrix}
		\frac{iq}{q_x-iq_y} \\
		1
	\end{matrix}\right) = -\frac{e^{-i\phi}}{\sqrt{2}} \left( \begin{matrix}
	1 \\
	-e^{i\phi}
\end{matrix}	 	\right),\nn
\hfill \nn
	&	\tilde{\Psi}_+(-q_x,\,q_y) = \frac{1}{\sqrt{2}} \left( \begin{matrix}
		-\frac{iq_x+q_y}{q} \\
		1
	\end{matrix}\right) = -\frac{e^{i\phi}}{\sqrt{2}} \left( \begin{matrix}
		1 \\
		-e^{-i\phi}
	\end{matrix}\right).
\end{align}
	\begin{figure}
		\centering
		\includegraphics[scale=0.3]{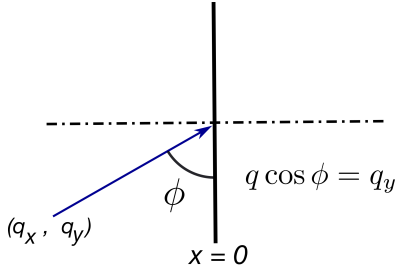}
		\caption{The incident angle as a function of the quasi momentum $\mbd{q}$ for the incident electrons.}
		\label{fig:Incidentangle}
	\end{figure}
The phase $\phi$ is defined in the way shown in Fig. \ref{fig:Incidentangle}. The total wavefunction on the left-hand side, denoted as $\Psi^{L}_+ (\mbd{r})$, can be expressed as a superposition of the incident and reflected wavefunctions, namely
\begin{align}
	\Psi^{L}_{+}(\mbd{r}) = e^{ik_x x+ik_y y} \tilde{\Psi}_+(q_x,\,q_y) + \mathcal{R} \, e^{-ik_x x+ik_y y} \, \tilde{\Psi}_+(-q_x,\,q_y).
\end{align}
The total wavefunction on the right-hand side $\Psi^{R}_+ (\mbd{r})$ is given by 
\begin{align}
	\Psi^{{R}}_{+}(\mbd{r}) = \mathcal{T} \, e^{ik_x x+ik_y y} \tilde{\Psi}_+(q_x,\,q_y)\end{align}
In the above, we neglect the phase factor and prefactor $1/\sqrt{2}$ since it will play no role on finding $\mathcal{R}$ and $\mathcal{T}$. Written $\eta = 1+ \delta t/t$, the B.C.'s can be found to be
\begin{align}
	\text{Boundary Conditions} \, \Rightarrow \begin{cases}
		\mathcal{T} e^{i\phi} = \eta \left( e^{i\phi}	 + \mathcal{R} \, e^{-i\phi} \right), \\
		1+\mathcal{R} = \eta \mathcal{T}.
	\end{cases}
\end{align}
We can solve for the transmission and reflection coefficients:
\begin{align}
	\mathcal{T} = \eta \, \frac{1-e^{-2i \phi}}{1-\eta^2 e^{-2i\phi}}~~;~~ \mathcal{R} = \frac{1-\eta^2}{\eta^2 e^{-2i\phi}-1}.
\end{align}
We can immediately check that once $\eta = 0 \, (\delta t = -t) $, there is no transmitted particles since the graphene has been cut into two separate pieces, and the probability current is conserved, that is $|\mathcal{R}|^2 + |\mathcal{T}|^2 = 1$.

\section{The boundary conditions for the Kane-Mele $d+id^\prime$ N/SC junction}
In this section, we derive the boundary condition for the electron scattering across the Kane-Mele $d+id^\prime$ -wave superconducting N/SC junction in the presence of a $\delta$-barrier for the $\mbd{K}_\pm$-valleys, respectively

%The wavefunction in momentum space is written as $\psi(q_{x},q_{y}) = (u_{A\uparrow},u_{B\uparrow},u_{A\downarrow},u_{B\downarrow},v_{A\uparrow},v_{B\uparrow},v_{A\downarrow},v_{B\downarrow})^{T}$, which satisfies the BdG equation
%\begin{align}
 % & \frac{3t}{2} (iq_{x}+q_{y})  u_{B\uparrow} +\frac{3 \Delta}{2} (-iq_{x}+q_{y}) v_{B\downarrow} -\delta t u_{B\uparrow} = E u_{A\uparrow}  \nn
  %& \frac{3t}{2} (-iq_{x}+q_{y})  u_{A\uparrow} +3 \Delta  v_{A\downarrow} -\delta t \, u_{A\uparrow} = Eu_{B\uparrow}. \nn
  %& \frac{3t}{2} (iq_{x}+q_{y})  u_{B\downarrow} -\frac{3 \Delta}{2} (-iq_{x}+q_{y}) v_{B\uparrow} - \delta t \, u_{B \downarrow} = Eu_{A\downarrow}.\nn
  %& \frac{3t}{2} (-iq_{x}+q_{y})  u_{A\downarrow} -3 \Delta  v_{A\uparrow} -\delta t \, u_{A\downarrow} = Eu_{B\downarrow}. \nn
  %& \frac{3t}{2} (-iq_{x}-q_{y})  v_{B\uparrow} -3 \Delta u_{B\downarrow} + \delta t v_{B\uparrow} = E v_{A\uparrow} \nn
   %&\frac{3t}{2} (iq_{x}-q_{y})  v_{A\uparrow} -\frac{3 \Delta}{2} (iq_{x}+q_{y}) u_{A\downarrow} + \delta t \, v_{A \uparrow} = Ev_{B\uparrow}.\nn
   %& \frac{3t}{2} (-iq_{x}-q_{y})  v_{B\downarrow} +3 \Delta u_{B\uparrow} + \delta t v_{B\downarrow} = E v_{A\downarrow} \nn
   %&\frac{3t}{2} (iq_{x}-q_{y})  v_{A\downarrow} +\frac{3 \Delta}{2} (iq_{x}+q_{y}) u_{A\uparrow} + \delta t \, v_{A \downarrow} = Ev_{B\downarrow}.
%\end{align}  

 %We can Fourier transform DBdG equations, and the  $\delta$ term coming from the local strain will generate a Delta-$\delta$ potential at the interface, $x = 0$.  
 The wavefunction for the DBdG Hamiltonian for the $\mbd{K}_-$-valley can be written as $\Psi(x, \, y) = (f_{A \uparrow}, \, f_{B \uparrow}, \, f_{A \downarrow}, \, f_{B \downarrow},\, g_{A \uparrow}, \, g_{B \uparrow}, \, g_{A \downarrow}, \, g_{B \downarrow} ) $. In real space, the DBdG equations with eigenenergy $\epsilon$ are given by
\begin{align}
     & \partial_{x} f_{B\uparrow}(x) -\frac{3 \Delta_0}{2 t}  \, \partial_{x} g_{B\downarrow}(x)  -\frac{\delta t}{t} \,\delta (x)  f_{B\uparrow} (x)= (\epsilon/t) f_{A\uparrow}(x),  \nn
    -  & \partial_{x}  f_{A\uparrow}(x) + \frac{3 \Delta_0}{2t}  g_{A\downarrow}(x) -\frac{\delta t}{t} \delta(x) \, f_{A\uparrow}(x) = (\epsilon/t) f_{B\uparrow}(x), \nn
     & \partial_{x}  f_{B\downarrow}(x) + \frac{3 \Delta_0}{2t} \, \partial_{x} g_{B\uparrow}(x) - \frac{\delta t}{t}  \delta(x) \, f_{B \downarrow} (x)=(\epsilon/t) f_{A\downarrow}(x),\nn
      -& \partial_{x}  f_{A\downarrow}(x) -\frac{\delta t }{t}\delta (x) \, f_{A\downarrow} (x)= (\epsilon/t) f_{B\downarrow}(x), \nn
      -  & \partial_{x}  g_{B\uparrow}(x) + \frac{\delta t}{t} \delta (x) g_{B\uparrow} (x)= (\epsilon/t) \, g_{A\uparrow}(x), \nn
         & \partial_{x}  g_{A\uparrow}(x) -\frac{3 \Delta_0	}{2t} \, \partial_{x} f_{A\downarrow} (x) + \frac{\delta t}{t} \delta (x) \, g_{A \uparrow}(x) = (\epsilon/t) \, g_{B\uparrow}(x),\nn
        - & \partial_{x}  g_{B \downarrow}(x) + \frac{\delta t }{t}\delta (x) g_{B \downarrow} (x)= (\epsilon/t) \, v_{A\downarrow}(x), \nn
       & \partial_{x}  g_{A\downarrow}(x) + \frac{3 \Delta_0}{2t} \, \partial_{x} f_{A \uparrow} (x) + \frac{\delta t}{t} \delta (x) \, g_{A \downarrow}(x) = (\epsilon/t) \, g_{B\downarrow}(x).
       \label{eq:BC-kmtj-1}
\end{align}

We integrate the above equations over a infinitesimal distance across the interface and Eq. (\ref{eq:BC-kmtj-1}) becomes
\begin{align}
    & \left[f_{B\uparrow}^{sc}(0) - f_{B\uparrow}^{N}(0) \right]-\frac{3 \Delta_0}{2t}  \, \left[ g_{B \downarrow}^{sc}(0) - g_{B \downarrow}^{N}(0)\right]  = \frac{\delta t}{t} f_{B\uparrow} (0) ,\nn
    -& \left[f_{A\uparrow}^{sc}(0) - f_{A\uparrow}^{N}(0)\right]  = \frac{\delta t}{t}  \, f_{A\uparrow}(0),\nn
   & \left[f_{B \downarrow}^{sc}(0)- f_{B \downarrow}^{N}(0) \right]+ \frac{3 \Delta_0}{2t} \, \left[g_{B\uparrow}^{sc}(0) - g_{B\uparrow}^{N}(0) \right] = \frac{\delta t}{t}\, f_{B \downarrow} (0),\nn
      - & \left[f_{A \downarrow}^{sc}(0)- f_{A \downarrow}^{N}(0) \right] = \frac{\delta t}{t}  \, f_{A\downarrow} (0), \nn
        & \left[g_{B \uparrow}^{sc}(0)- g_{B \uparrow}^{N}(0) \right] = \frac{\delta t}{t}  g_{B\uparrow} (0), \nn
       & \left[g_{A \uparrow}^{sc}(0)- g_{A \uparrow}^{N}(0) \right] -\frac{3 \Delta_0}{2t} \, \left[f_{A \downarrow}^{sc}(0)- f_{A \downarrow}^{N}(0) \right]  = - \frac{\delta t}{t} \, g_{A \uparrow}(0),\nn
         & \left[g_{B \downarrow}^{sc}(0)- g_{B \downarrow}^{N}(0) \right]  =  \frac{\delta t}{t} g_{B \downarrow} (0), \nn
        & \left[g_{A \downarrow}^{sc}(0)- g_{A \downarrow}^{N}(0) \right]  + \frac{3 \Delta_0}{2t} \, \left[f_{A \uparrow}^{sc}(0)- f_{A \uparrow}^{N}(0) \right]   = \frac{\delta t}{t}  \, g_{A \downarrow}(0).
        \label{eq: BC-integrate}
\end{align}
In the above equations for the boundary conditions, $f_{\alpha \sigma} (0^{+}) \equiv f_{\alpha \sigma} ^{sc}$ while $f_{\alpha \sigma} (0^{-}) \equiv f_{\alpha \sigma} ^{N}$ with $\alpha$ denotes the $A, \, B$ sublattice and $\sigma$ denotes the spin $\sigma = \uparrow, \, \downarrow$. Similarly, the wavefunction ambiguity at $x = 0$ also exists in Eq, (\ref{eq: BC-integrate}) due to the fact that the DBdG equations are of first order. Following the similar approach, we are able to  determine $\Psi(0)$ via the requirements of no tunnelling current across the N/SC junction as $\Delta_0 = 0$ and $\delta t = -t$. Based on Eq. (\ref{eq:BC-ambiguity}), the boundary conditions at $x=0$ can be written as
\begin{align}
   f_{B \uparrow}^{sc}(0) & = \eta \, f_{B \uparrow}^{N}(0)  +  \frac{ 3\Delta_0}{2t} \, \left[ g_{B \downarrow}^{sc}(0) - g_{B \downarrow}^{N}(0)\right], \nn
     f_{A \uparrow}^{N}(0) & = \eta \,f_{A \uparrow}^{sc}(0) ,  \nn
    f_{B \downarrow}^{sc}(0) & = \eta \, f_{B \downarrow}^{N}(0)  - \frac{ 3 \Delta_0}{2t} \, \left[ g_{B \uparrow}^{sc}(0) - g_{B \uparrow}^{N}(0)\right]  , \nn
      f_{A \downarrow}^{N}(0) & = \eta \, f_{A \downarrow}^{sc}(0) , \nn
       g_{B \uparrow}^{sc}(0) & = \eta \, g_{B \uparrow}^{N}(0), \nn
       g_{A \uparrow}^{N}(0) & =\eta \,g_{A \uparrow}^{sc}(0)  -  \frac{3 \Delta_0}{2t} \, \left[ f_{A \downarrow}^{sc}(0) - f_{A \downarrow}^{N}(0)\right], \nn
         g_{B \downarrow}^{sc}(0) & = \eta \, g_{B \downarrow}^{N}(0), \nn
         g_{A \downarrow}^{N}(0) & = \eta \, g_{A \downarrow}^{sc}(0)  +  \frac{ 3\Delta_0}{2t} \, \left[ f_{A \uparrow}^{sc}(0) - f_{A \uparrow}^{N}(0)\right],
        \label{eq: BC-final}
\end{align}
where $\eta \equiv  1 + \delta t/t$. Following the same procedures, the boundary condition for $\mbd{K}_+$-valley are given by
\begin{align}
	 f_{B \uparrow}^{sc}(0) & = \eta \, f_{B \uparrow}^{N}(0), \nn
     f_{A \uparrow}^{N}(0) & = \eta \,f_{A \uparrow}^{sc}(0)-  \frac{ 3 \Delta_0}{2t} \,  \left[g^{sc}_{A \downarrow} (0)-g^N_{A \downarrow}(0)\right],  \nn
    f_{B \downarrow}^{sc}(0) & = \eta \, f_{B \downarrow}^{N}(0)  , \nn
      f_{A \downarrow}^{N}(0) & = \eta \, f_{A \downarrow}^{sc}(0)  + \frac{ 3 \Delta_0}{2t} \, \left[ g_{A \uparrow}^{sc}(0) - g_{A \uparrow}^{N}(0)\right] , \nn
       g_{B \uparrow}^{sc}(0) & = \eta \, g_{B \uparrow}^{N}(0) + \frac{ 3 \Delta_0}{2t} \, \left[ f_{B \downarrow}^{sc}(0) - f_{B\downarrow}^{N}(0)\right] , \nn
       g_{A \uparrow}^{N}(0) & =\eta \,g_{A \uparrow}^{sc}(0) , \nn
         g_{B \downarrow}^{sc}(0) & = \eta \, g_{B \downarrow}^{N}(0)  -  \frac{3 \Delta_0}{2t} \, \left[ f_{B \uparrow}^{sc}(0) - f_{B \uparrow}^{N}(0)\right], \nn
         g_{A \downarrow}^{N}(0) & = \eta \, g_{A \downarrow}^{sc}(0).  
        \label{eq: BC-final-2}
\end{align}

%%%%%%%%%%%%%%%%%%%
%%%%%%%%%%%%%%%%%%%

\end{document}